\documentclass[iop, revtex4, numberedappendix, twocolappendix]{emulateapj}

\usepackage[backref,breaklinks,colorlinks,citecolor=blue]{hyperref}
\usepackage{epstopdf}
\usepackage{amssymb}
\usepackage{amsmath}
\usepackage[pdftex]{lscape}
\usepackage[pass]{geometry}
\usepackage[all]{hypcap}
\usepackage[usenames, dvipsnames]{color}
\usepackage{ulem}
\def\lta{{\>\rlap{\raise2pt\hbox{$<$}}\lower3pt\hbox{$\sim$}\>}}
\def\gta{{\>\rlap{\raise2pt\hbox{$>$}}\lower3pt\hbox{$\sim$}\>}}

\shorttitle{Dust attenuation, bulge formation and inside-out quenching of star-formation at $z\sim2$}
\shortauthors{Tacchella, Carollo, F\"orster Schreiber et al.}

\begin{document}
\title{Dust attenuation, bulge formation and inside-out quenching of star-formation  \\ in Star-Forming Main Sequence galaxies at $\MakeLowercase{z}\sim2$ \altaffilmark{\dag}}

\author{S. Tacchella\altaffilmark{1,2}, 
 C. M. Carollo\altaffilmark{1},
 N. M. F\"orster Schreiber\altaffilmark{3},
 A. Renzini\altaffilmark{4,5},
 A. Dekel\altaffilmark{6},
 R. Genzel\altaffilmark{3,7,8},
 P. Lang\altaffilmark{3}, \\
 S. J. Lilly\altaffilmark{1},
 C. Mancini\altaffilmark{4},
 M. Onodera\altaffilmark{9,10},
 L. J. Tacconi\altaffilmark{3},
 S. Wuyts\altaffilmark{11},
 G. Zamorani\altaffilmark{12}
}

\altaffiltext{1}{Department of Physics, Institute for Astronomy, ETH Zurich, CH-8093 Zurich, Switzerland}
\altaffiltext{2}{Harvard-Smithsonian Center for Astrophysics, 60 Garden St., Cambridge, MA 02138, USA}
\email{sandro.tacchella@cfa.harvard.edu} 
\altaffiltext{3}{Max-Planck-Institut f\"ur extraterrestrische Physik (MPE), Giessenbachstr. 1, D-85748 Garching, Germany}
\altaffiltext{4}{INAF Osservatorio Astronomico di Padova, vicolo dell’Osservatorio 5, I-35122 Padova, Italy}
\altaffiltext{5}{Department of Physics and Astronomy Galileo Galilei, Universita degli Studi di Padova, via Marzolo 8, I-35131 Padova, Italy}
\altaffiltext{6}{Center for Astrophysics and Planetary Science, Racah Institute of Physics, The Hebrew University, Jerusalem 91904, Israel}
\altaffiltext{7}{Department of Astronomy, Campbell Hall, University of California, Berkeley, CA 94720, USA}
\altaffiltext{8}{Department of Physics, Le Conte Hall, University of California, Berkeley, CA 94720, USA}
\altaffiltext{9}{Subaru Telescope, National Astronomical Observatory of Japan, National Institutes of Natural Sciences (NINS), 650 North A'ohoku Place, Hilo, HI 96720, USA}
\altaffiltext{10}{Graduate University for Advanced Studies, 2-21-1 Osawa, Mitaka, Tokyo, Japan}
\altaffiltext{11}{Department of Physics, University of Bath, Claverton Down, Bath, BA2 7AY, UK}
\altaffiltext{12}{INAF Osservatorio Astronomico di Bologna, Via Ranzani 1, I-40127 Bologna, Italy}

\altaffiltext{\dag}{Based on observations made with the NASA/ESA {\em Hubble Space Telescope\/}, obtained at the Space Telescope Science Institute, which is operated under NASA contract NAS 5$-$26555 (programs GO9822, GO10092, GO10924, GO11694, GO12578, GO12060, GO12061, GO12062, GO12063, GO12064, GO12440, GO12442, GO12443, GO12444, GO12445, GO13669), and at the {\em Very Large Telescope\/} of the European Southern Observatory, Paranal, Chile (ESO Programme IDs 075.A-0466, 076.A-0527, 079.A-0341, 080.A-0330, 080.A-0339, 080.A-0635, 081.A-0672, 183.A-0781, 087.A-0081, 088.A-0202, 088.A-0209, 091.A-0126).}

\slugcomment{{\sc Draft Version:} \today}

\begin{abstract}
We derive 2D dust attenuation maps at $\sim1~\mathrm{kpc}$ resolution from the UV continuum for 10 galaxies on the $z\sim2$ star-forming main sequence (SFMS). Comparison with IR data shows that 9 out of 10 galaxies do not require further obscuration in addition to the UV-based correction, though our sample does not include the most heavily obscured, massive galaxies. The individual rest-frame $V$-band dust attenuation (A$_{\rm V}$) radial profiles scatter around an average profile that gently decreases from $\sim1.8$ mag in the center down to $\sim0.6$ mag at $\sim3-4$ half-mass radii. We use these maps to correct UV- and H$\alpha$-based star-formation rates (SFRs), which agree with each other. At masses $\lta10^{11}~M_{\odot}$, the dust-corrected specific SFR (sSFR) profiles are on average radially constant at a mass-doubling timescale of $\sim300~\mathrm{Myr}$, pointing at a synchronous growth of bulge and disk components. At masses $\gta10^{11}~M_{\odot}$, the sSFR profiles are typically centrally suppressed by a factor of $\sim10$ relative to the galaxy outskirts. With total central obscuration disfavored, this indicates that at least a fraction of massive $z\sim2$ SFMS galaxies have started their inside-out star-formation quenching that will move them to the quenched sequence. In combination with other observations, galaxies above and below the ridge of the SFMS relation have respectively centrally enhanced and centrally suppressed sSFRs relative to their outskirts, supporting a picture where bulges are built owing to gas `compaction' that leads to a high central SFR as galaxies move toward the upper envelope of the SFMS. \\
 \end{abstract}


\section{Introduction} \label{sec:intro}

The existence at any epoch of an almost linear correlation between star-formation rate (SFR) and stellar mass ($M_{\star}$), i.e. the star-torming main sequence (SFMS; \citealt{brinchmann04, daddi07, noeske07, salim07, rodighiero11, whitaker12, whitaker14, speagle14, rodighiero14, schreiber15}), suggests that galaxies grow in mass and size with cosmic time in a state of self-regulated semi-equilibrium \citep[e.g.,][]{daddi10, bouche10, genzel10, tacconi10, dave12, lilly13_bathtube, dekel13, dayal13, feldmann15, tacchella16_MS}. Understanding the details of this equilibrium, as well as the processes that permanently move galaxies out of the SFMS onto the `quenched' population\footnote{We refer to quenched galaxies as galaxies that do not double their stellar mass within the Hubble time, i.e. $\mathrm{sSFR}^{-1}>t_{\rm H}$.}, necessitates spatially resolved information within individual galaxies of their stellar mass and SFR density distributions. This is particularly important at redshifts of order $z\sim2$, the epoch of the peak of the cosmic SFR density and of the assembly of a large fraction of the stellar mass that is seen locked in the $z=0$ massive spheroidal population. 

Our SINS/zC-SINF program of Very Large Telescope (VLT) adaptive optics (AO) SINFONI integral field spectroscopy and Hubble Space Telescope (\textit{HST}) imaging has returned a wealth of facts on galaxies on the $z\sim2$ SFMS (see Section~\ref{subsec:Sample} for details and references). Specifically, in \citet{tacchella15_sci} we constrained the $M_{\star}$ and SFR distributions resolved on $\sim1$ kpc scales in $\sim30$ such galaxies with stellar masses above $\sim10^{9.5}~M_\odot$. We found that, at the lower-mass end of our sample, $M_{\star}\lta10^{11}~M_{\odot}$, galaxies have flat specific SFR (sSFR) profiles on average, indicating that they are doubling their mass at all radii with the same pace. In contrast, at masses of $M_{\star}\sim10^{11}~M_{\odot}$ and slightly above, the sSFR profiles decrease toward the galaxy centers to values of $\mathrm{sSFR}^{-1}\ga~t_{\rm H}$ (with $t_{\rm H}$ the Hubble time), suggesting that these galaxies have started their descent toward the quenched population by decreasing their star-formation activity in their centers (i.e. quenching `inside-out'). In addition, the central mass density in such massive star-forming galaxies appears to have already reached the high values that characterize the $z=0$ quenched spheroidal population of similar mass, consistent with results by \citet{van-dokkum10, van-dokkum14_dense_cores} and \citet{saracco12}. 

These results carry implications for both spheroid formation and quenching mechanisms at those early epochs. \citet{genzel14a} measured large nuclear Toomre $Q$-values in the same massive galaxies, which they interpreted as indicating that suppression of giant clump formation is responsible for the centrally suppressed SFRs in such gas-rich high-$z$ disks. More generally, the presence of spheroid-like stellar densities in massive SFMS galaxies with quasi-quenched cores argues for a direct link between SFMS progenitors and quenched descendants, and in turn for a continuous feeding of the quenched population with galaxies whose sizes increase with cosmic time following the same scaling of the star-forming population. This `progenitor bias' effect \citep{van-dokkum96} has indeed been argued in some works to be the driver of most of the observed growth of the average size of the quenched population at masses of order $\sim10^{11}~M_\odot$ \citep{valentinuzzi10, saracco11, carollo13a, cassata13, poggianti13, fagioli16, williams17}, i.e. the SFR- and epoch-independent characteristic mass of the galaxy mass function since at least $z\sim4$ \citep[e.g.,][]{ilbert13, muzzin14}, which at any epoch entails the bulk of the spheroidal population. Only a small portion ($\sim15\%$) of the $z=0$ spheroid population reaches masses substantially above this characteristic mass; the structural and kinematic properties of such ultramassive and rare spheroids show unequivocal evidence for a dissipationless formation history \citep[][and references therein]{bender89, carollo93, faber97, binney08, emsellem11, cappellari16}.

Analytical and numerical calculations provide a benchmark for interpreting the observed mass and sSFR profiles. \citet{lilly16} show that a mass-dependent quenching mechanism such as in, e.g., \citet{peng10_Cont}, acting on star-forming disks whose sSFRs and sizes follow the cosmic evolution of these parameters, leads to a stratification of stellar density in SFMS galaxies that indeed achieves spheroidal densities at the onset of quenching (without, however, any causal connection between stellar density and quenching, which, in the model, is entirely driven by total stellar mass). In \citet{tacchella16_profile} and \citet{tacchella16_MS} we examined the VELA cosmological zoom-in simulations of \citealt{ceverino14_radfeed} (see also \citealt{zolotov15}) and found that profiles of stellar mass and SFR such as those reported in \citet{tacchella15_sci} are realized through up-and-down oscillation cycles within the upper and lower boundaries of the SFMS. The physical reason behind these oscillations is the alternate occurrence of strong inward gas flows and gas depletion through gas consumption and outflows (driven by feedback). The strong inward gas flows lead to substantial compression of the gas reservoir in the galaxy centers (a process that we refer to as `compaction', see \citealt{dekel14_nugget, zolotov15, tacchella16_profile}). Compaction leads to strong central starbursts that push galaxies toward the upper SFMS envelope and add stellar mass to the bulge components. This compaction event is followed by gas depletion through gas consumption and outflows (driven by feedback), which pushes galaxies down toward the lower envelope of the SFMS.

An important issue, however, remains: to establish how the observed shapes of the sSFR profiles are affected by dust extinction. More generally, the spatially resolved dust attenuation distribution in high-$z$ galaxies still is poorly understood owing to the scarcity of empirical constraints. However, it can have a significant impact not only on the inferred star formation distribution of galaxies but also on the measurement of sizes and shapes \citep[e.g.,][]{van-der-wel14, van-der-wel14a, tacchella15}, the estimation of the stellar mass surface density \citep[e.g.,][]{wuyts12, lang14, tacchella15_sci}, the identification of star forming clumps \citep[e.g.,][]{cibinel17, guo18}, and the conversion of the H$\alpha$ luminosity to the gas surface density \citep[e.g.,][]{genzel14a}, to name a few examples. It is therefore important to investigate the dust attenuation distribution and study its impact on other measured quantities. We will focus here in this paper specifically on the impact of the dust attenuation on the inferred star-formation distribution. In \citet{tacchella15_sci} we have adopted a uniform attenuation A$_{\rm H\alpha}$ over the face of galaxies. We found that in the most massive galaxies ($M_{\star}>10^{11}~M_{\odot}$) the sSFR is substantially depressed at the center while leveling off to high values toward the outskirts. Yet, the assumption of a uniform attenuation was crucial, a limitation that we try to alleviate with this paper. Here we use $HST$ $B$- and $I$-band imaging (with $\sim1$ kpc resolution) to construct spatially resolved, rest-frame (FUV$-$NUV) color maps from which we derive the UV continuum slope ($\beta$) and from it the UV attenuation.

Generally, the main approach to correct for dust attenuation relies on applying a wavelength-dependent dust attenuation curve \citep[e.g.,][]{seaton79, cardelli89, fitzpatrick99, reddy15} to observational constraints such as the Balmer decrement \citep[e.g.,][]{calzetti97}, the ratio of far-infrared (far-IR) to ultraviolet (UV) emission \citep[$\mathrm{IRX}=L_{\rm IR}/L_{\rm UV}$, e.g.,][]{buat96, witt00, panuzzo03, buat05}, or the UV continuum slope \citep{calzetti94, meurer99}. The most reliable technique to estimate the dust attenuation is to measure the  flux ratio of two nebular Balmer emission lines such as H$\alpha$/H$\beta$ (i.e., the Balmer decrement). Since the value of the Balmer decrement is set by quantum physics, any deviation from this expected value may be attributed to dust extinction (for a fixed electron temperature). Moreover, if dust attenuation is highly patched, this ratio may just reflect the less  attenuated regions of a galaxy. However, the simultaneous detection of H$\alpha$/H$\beta$ in higher-redshift galaxies is observationally challenging, in particular on spatially resolved scales. Stacking data of several hundred galaxies from the 3D-HST survey \citep{brammer12, skelton14, momcheva16}, \citet{nelson16_balmer} derived Balmer decrement maps in $z\sim1.4$ galaxies, finding A$_{\rm H\alpha}\approx3$ mag of dust attenuation localized within the innermost $\approx1~\mathrm{kpc}$.

With the vastly improved sensitivity at submillimeter and millimeter wavelengths provided by ALMA and NOEMA, it is now possible to measure on spatially resolved scales the obscured star formation at $z>1$. To date, only a small number of high-$z$ galaxies have been studied on spatially resolved scales \citep[e.g.,][]{tadaki15, rujopakarn16, barro16, tadaki17, cibinel17, nelson18}. Some of these studies \citep{tadaki15, barro16, tadaki17} find of the order of $3-5~\mathrm{mag}$ extra attenuation in the galaxy centers relative to their outskirts. However, the selection criteria for those samples (i.e. compact sizes of $\sim1-2$ kpc, high IR fluxes and/or very red colors) are clearly different from those for typical SFMS galaxies that have sizes of $1-5$ kpc.

Substantially shallower dust attenuation radial gradients are reported in larger samples of high-$z$ star-forming galaxies (SFGs) by \citet{wuyts12}, \citet{hemmati15} and \citet{wang17_dust}, based on a pixel-by-pixel spectral energy distribution (SED) modelling of photometric data in CANDELS, where the dust attenuation is constrained mainly from the UV color. The UV continuum slope has been traditionally used to do the dust correction by relating the the UV slope $\beta$ to the dust attenuation in the UV (i.e. A$_{\rm UV}=4.43+1.99\beta$, \citealt{meurer99}). This relation is itself derived from the IRX$-\beta$ relation. Under the assumption that all star-forming galaxies / regions have similar intrinsic UV slopes, and that their IR luminosity arises from dust heated by the same UV continuum, it is possible to show that there is a unique relation between IRX and A$_{\rm UV}$. The great utility of this relation is then that it allows dust-corrected SFRs to be derived based on nothing more than an apparently straightforward measurement of the UV luminosity and the spectral slope. Thus, estimating SFRs from UV luminosities corrected for attenuation in this way has become common practice especially in high-redshift studies. Still, this is certainly an imperfect way of constructing sSFR maps for high-redshift galaxies. Indeed, we know that for a minority of starburst galaxies this method dramatically underestimates the actual SFR as instead traced by the FIR luminosity \citep[e.g.,][]{rodighiero11}. However, on average the SFRs from attenuation-corrected UV luminosities appear to agree with FIR-based ones, as demonstrated by stacking \textit{Herschel} data for SFMS galaxies at $z\sim 2$ \citep{rodighiero14}.

Still, in the local universe, it has been shown that UV continuum slope is poorly correlated with attenuation, as probed by IRX, particularly on spatially resolved scales \citep[e.g.,][]{mao12, boquien12, hao11}. The main cause for this is that the key assumptions are breaking down: individual star-forming regions can have significantly different ages and hence different intrinsic UV slopes. Additionally, on small subkiloparsec spatial scales, as probed by some of these $z\sim0$ data, one expects that the IR luminosity at a certain position is heated by stars at a different position. At higher redshifts, these concerns will be partially alleviated since the age spread in the stellar population is smaller owing to the upper limit set by the age of the universe and the spatial resolution of the data is lower, but this must be tested and the spread in the IRX$-\beta$ relation quantified. Testing the IRX$-\beta$ relationship at high redshift has been the matter of several studies using \textit{Spitzer}, \textit{Herschel}, and recently also in combination with ALMA observations \citep[e.g.,][]{reddy10, reddy12a, reddy18, nordon13, mclure18, koprowski18}.  For typical but little or modestly obscured systems, the results are broadly consistent with the \citealt{meurer99} IRX$-\beta$ relationship (and strongly deviate from it for more IR-luminous, highly obscured galaxies), although there is still a debate on whether they support a gray attenuation curve \citep[e.g.,][]{calzetti00, mclure18, koprowski18} or a more SMC-like one \citep[e.g.][]{reddy18}. On spatially resolved scales, only little progress has been made at high $z$. A recent study by \citet{nelson18} shows that, at least for one galaxy at $z\sim1.2$ (with $M_{\star}\approx6.8\times10^{10}~M_{\odot}$ and $\mathrm{SFR}\approx170~M_{\odot}/$yr), the dust-corrected UV SFR profile agrees very well with the IR SFR profile, using for the dust correction the dust attenuation estimate from the UV-optical SED, indicating that the dust-corrected rest-frame UV SFRs are not substantially missing light from regions of very high dust obscuration.

With all these provisos, in this paper we assume that SFRs derived from the attenuation-corrected UV luminosity are reliable in all but possibly a minority of cases with extremely high attenuation, with A$_{\rm UV}$ being derived from the UV slope $\beta$ as mentioned above. Moreover, we shall also assume that the UV attenuation A$_{\rm UV}$ can be used to derive the attenuation at H$\alpha$ (A$_{\rm H\alpha}$), hence allowing us to construct space-resolved SFR maps from locally corrected H$\alpha$ flux maps. Specifically, we constrain the spatial distribution of the dust attenuation and its impact on the measured star-formation distribution in our SINS/zC-SINF sample. We use new Cycle 22 \textit{HST} F438W ($B$) and F814W ($I$) imaging data (\# GO13669). These trace respectively the far-ultraviolet (FUV) and near-ultraviolet (NUV) light distributions within our $z\sim2$ galaxies, which enables us to measure maps of dust attenuation from the UV continuum slope on resolved on scales of $\sim1~\mathrm{kpc}$. In the light of all the caveats and uncertainties mentioned above, we will show that the dust-corrected UV SFRs are in good agreement with IR+UV SFRs, demonstrating that, in our sample, the rest-frame UV and H$\alpha$ SFRs are not substantially missing light from regions of very high dust obscuration. While our dust attenuation profiles steadily increase toward the centers, they do so with a relatively shallow slope, which results in a significant dust attenuation of $\mathrm{A}_{\rm V}\approx0.6~\mathrm{mag}$ out to $\sim10~\mathrm{kpc}$. Importantly, the centrally suppressed sSFR in our massive sample cannot be explained by dust attenuation alone and is evidence of a genuine reduction of star-formation activity in the centers of massive galaxies on the SFMS at $z \sim 2$.

The paper is organized as follows. In Section~\ref{sec:Sample_Data}, we present the galaxy sample and the data. We review the methodology for deriving dust attenuation and SFR diagnostics in SFGs in Section~\ref{sec:method} and present the resulting measurements in Section~\ref{sec:measurement}. In particular, in Section~\ref{subsec:SFR_Profiles} we quantify the impact on the SFR density profiles of assuming different dust attenuation corrections. In Section~\ref{sec:discussion} we discuss where they are sustaining the bulk of their star-formation activity and the growth in stellar mass, i.e. whether in their bulge or outskirt regions. We summarize in Section~\ref{sec:summary}.

Throughout this paper, we adopt WMAP9 cosmology: $H_0=69.3~\mathrm{km~s^{-1}~Mpc^{-1}}$, $\Omega_{\Lambda,0}=0.71$, and $\Omega_{m,0}=0.29$ \citep{hinshaw13}. For this cosmology, $1\arcsec$ corresponds to $\approx8.4~\mathrm{kpc}$ at $z=2.2$. All sizes and radii presented in this paper are circularized, i.e. $r=r_a\sqrt{(b/a)}$. Magnitudes are given in the AB photometric system.

\section{Sample and Data}\label{sec:Sample_Data}

\subsection{Galaxy Sample} \label{subsec:Sample}

\begin{figure*}
\begin{center}
\includegraphics[width=0.85\textwidth]{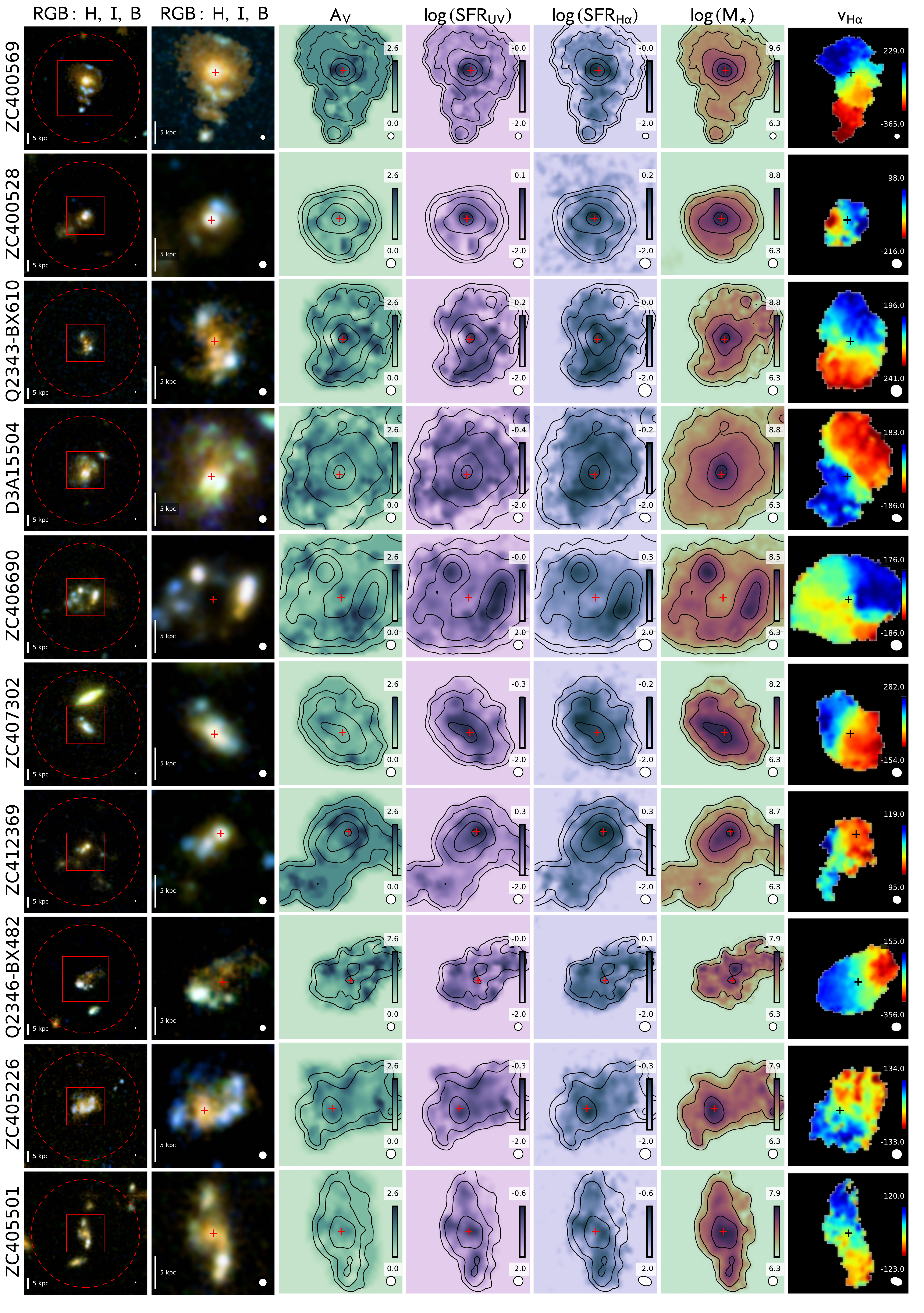} 
\caption{From left to right: \textit{HST} RGB image (red: $H$, green: $I$, blue: $B$, all observed frame), A$_{\rm V}$ dust attenuation map (in mag), dust-corrected UV SFR map (in $M_{\odot}~\mathrm{yr}^{-1}~\mathrm{kpc}^{-2}$), dust-corrected H$\alpha$ SFR map (in $M_{\odot}~\mathrm{yr}^{-1}~\mathrm{kpc}^{-2}$), stellar mass map (in $M_{\odot}~\mathrm{kpc}^{-2}$), and H$\alpha$ velocity map (in $\mathrm{km}~\mathrm{s}^{-1}$). Red boxes in the leftmost images show the field of view of the SINFONI H$\alpha$ maps, and the red dashed circle indicates the 3\arcsec aperture of the photometry. The rulers in the bottom left of the maps indicate 5 kpc and the circles in the bottom right show the size of the PSF. The contours indicate the stellar mass surface density between $\log\Sigma_{\rm M}/(M_{\odot}~\mathrm{kpc}^2)=6.5$ and 9.0 in linear steps of 0.5. } 
\label{fig:Maps}
\end{center}
\end{figure*}

\begin{figure*}
\includegraphics[width=0.95\textwidth]{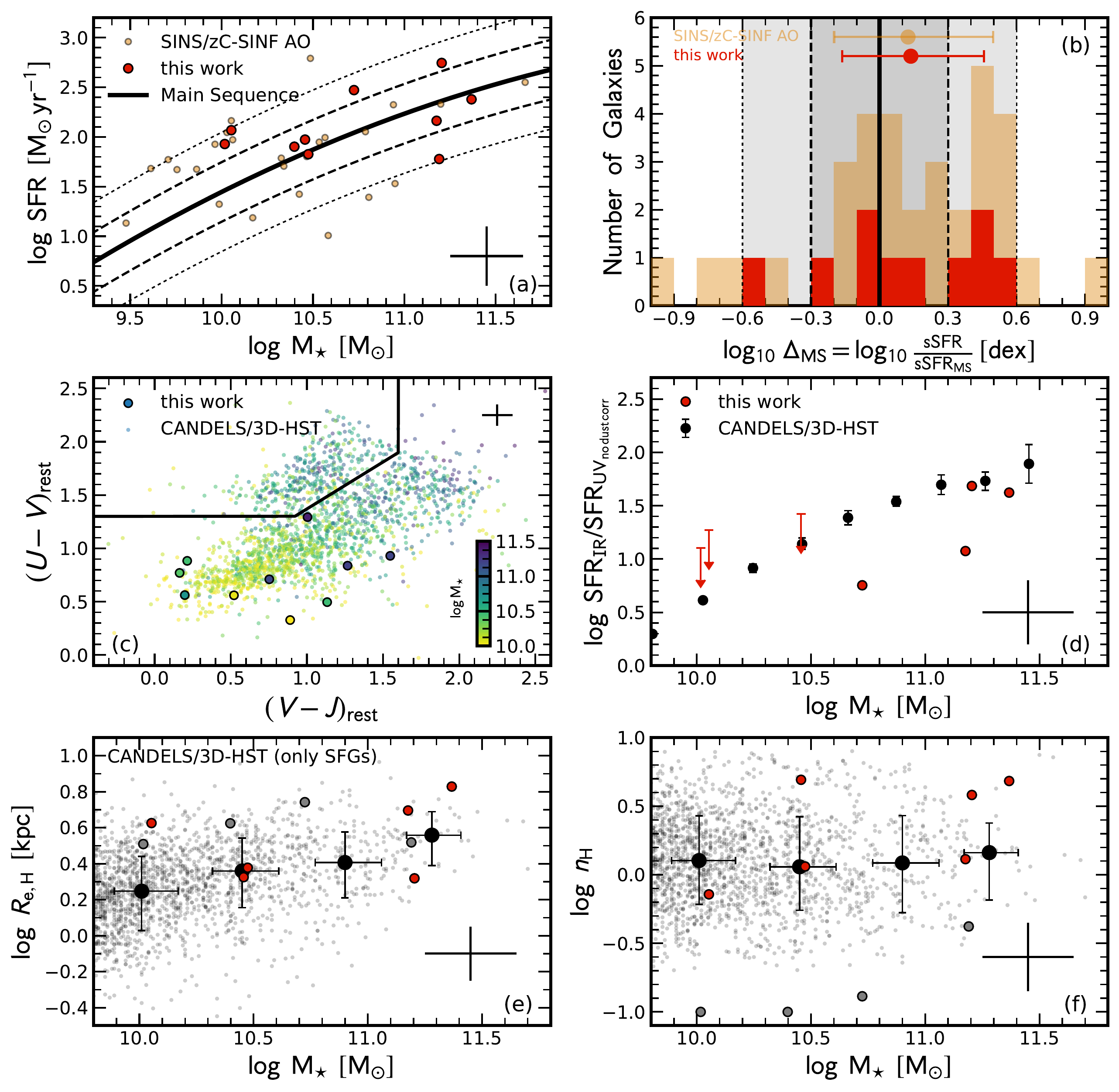} 
\caption{Our sample of galaxies relative to the global population at $z=2.0-2.5$. The stellar mass is defined to be the integral of the past SFR; all results from the literature are adjusted to this definition. \textit{Panel (a):} The 10 galaxies of this study (shown with red circles) and the parent sample SINS/zC-SINF AO (orange circles) in the SFR$-M_{\star}$ plane. The quoted SFRs are UV+IR SFR (if available and reliable; otherwise SED-based SFRs). The thick solid line shows the main-sequence ridgeline of \citet{whitaker14} at $z=2.2$ (median redshift our sample), which is based on the CANDELS/3D-HST survey. The dashed and dotted lines indicate the $1\sigma$ and $2\sigma$ scatter of the Main Sequence. All 10 galaxies of our sample lie within $2\sigma$ of the main sequence. \textit{Panel (b):} histogram of the distance from the main-sequence ridgeline $\Delta_{\rm MS}$. Our galaxies are well distributed around the main-sequence ridgeline. The median $\Delta_{\rm MS}$ for our sample and the parent sample are $0.10^{+0.38}_{-0.28}$ and $0.14^{+0.32}_{-0.30}$, respectively. \textit{Panel (c):} Our sample in the UVJ rest-frame color-color diagram. The circles show the global $z=2.0-2.5$ galaxy population with $M_{\star}=10^{10.0}-10^{11.5}~M_{\odot}$ drawn from the CANDELS/3D-HST survey, color-coded by their stellar mass. While our sample overlaps with global population, we do not probe the upper right region. \textit{Panel (d):} Ratio SFR$_{\rm IR}$/SFR$_{\rm UV}$ as a function of $M_{\star}$. Note that the UV SFR is not corrected for extinction. The black filled circles show the median relation of \citet{whitaker14} at $z=2.0-2.5$, where the IR data was stacked. Several galaxies of our sample are not detected in the IR (plotted are upper limits) or have no IR coverage (are not plotted; see also Figure~\ref{fig:SFR_UV_vs_IR} below). Overall, our galaxies probe the trend of the population, but only four galaxies have a reliable IR detection. \textit{Panel (e) and (f):} Circularized $H$-band half-light radius $R_{\rm e, H}$ and S\'{e}rsic index $n_{\rm H}$ as a function of the total stellar mass $M_{\star}$. The gray circles indicate unreliable surface brightness fits because light distribution is not centrally concentrated (assumed S\'{e}sic models do not represent the galaxy well). The small transparent gray points show the measurements of \citet{van-der-wel14a} at $z=2.0-2.5$ for star-forming galaxies ($UVJ$-selected), while the black filled circles show their median relation. Overall, our galaxies follow the main trends of the global galaxy population in half-light radius and S\'{e}rsic index.} 
\label{fig:Sample}
\end{figure*}

\capstartfalse 
\begin{deluxetable*}{lcccccccccc}
\tabletypesize{\scriptsize}
\tablecolumns{11}
\tablewidth{0pc}
\tablecaption{Sample galaxies with H$\alpha$ redshifts and main stellar population properties. \label{tbl:sample}}
\tablehead{\colhead{Source} & \colhead{$z_{\mathrm{H\alpha}}$} & \colhead{M$_{\star}$} & \colhead{$(U-V)_{\rm rest}$} & \colhead{A$_{\rm V, SED}$} & \colhead{A$_{\rm V, map}$} & \colhead{SFR$_{\rm SED}$} & \colhead{SFR$_{\rm UV+IR}$} & \colhead{Indicator} & \colhead{SFR$_{\rm UV}$} & \colhead{SFR$_{\rm UV, map}$} \\ 
\colhead{} & \colhead{} & \colhead{[10$^{10}$ M$_{\sun}$]} & \colhead{[mag]} & \colhead{[mag]} & \colhead{[mag]} & \colhead{[$\frac{\mathrm{M}_{\sun}}{yr}$]} & \colhead{[$\frac{\mathrm{M}_{\sun}}{yr}$]} & \colhead{} & \colhead{[$\frac{\mathrm{M}_{\sun}}{yr}$]} & \colhead{[$\frac{\mathrm{M}_{\sun}}{yr}$]}}
\startdata
ZC400569 & 2.241 & 23.3 & 1.29 & 1.4 & 1.5 & 241.0 & 239.3 & UV+24$\mu$m & 168.0 & 156.6 \\
ZC400528 & 2.387 & 16.0 & 0.84 & 0.9 & 0.5 & 148.0 & 556.5 & UV+100$\mu$m & 148.0 & 46.7 \\
Q2343-BX610 & 2.211 & 15.5 & 0.93 & 0.8 & 1.2 & 60.0 & --- & --- & 60.0 & 85.5 \\
D3A15504 & 2.383 & 15.0 & 0.71 & 1.0 & 1.0 & 150.2 & 145.6 & UV+24$\mu$m & 150.0 & 116.1 \\
ZC406690 & 2.195 & 5.3 & 0.56 & 0.7 & 1.0 & 200.0 & 296.5 & UV+24$\mu$m & 337.0 & 201.2 \\
ZC407302 & 2.182 & 2.98 & 0.50 & 1.3 & 0.6 & 340.0 & 358.1$^\dagger$ & UV+24$\mu$m & 130.0 & 67.0 \\
ZC412369 & 2.028 & 2.86 & 0.88 & 1.0 & 1.3 & 94.1 & IR-undet & --- & 130.0 & 140.2 \\
Q2346-BX482 & 2.257 & 2.5 & 0.77 & 0.8 & 1.1 & 79.8 & --- & --- & 80.0 & 95.8 \\
ZC405226 & 2.287 & 1.13 & 0.56 & 1.0 & 0.9 & 117.0 & IR-undet & --- & 87.0 & 58.3 \\
ZC405501 & 2.154 & 1.04 & 0.33 & 0.9 & 0.6 & 84.9 & IR-undet & --- & 68.0 & 28.1
\enddata
\tablecomments{Listed are the H$\alpha$ spectroscopic redshifts from the AO-SINFONI data ($z_{\rm H\alpha}$), the stellar masses ($M_{\star}$; defined as the integral of the SFR), the rest-frame $U-V$ colors, the dust attenuation A$_{\rm V, SED}$ from galaxy-integrated SED modeling, the dust attenuation A$_{\rm V, map}$ from the $(\mathrm{FUV}-\mathrm{NUV})$ color maps, the SFRs from SED (SFR$_{\rm SED}$), the UV+IR SFRs (SFR$_{\rm UV+IR}$), the SFR indicator of the IR, SFR from the aperture UV photometry (SFR$_{\rm UV}$), and the SFR from the UV maps (SFR$_{\rm UV, map}$). For the SED modeling we use the \citet{bruzual03} model and assume a \citet{chabrier03} IMF, solar metallicity, the \citet{calzetti00} reddening law, and either constant or exponentially declining SFRs. The uncertainties on the stellar properties are dominated by systematics from the model assumption and are up to a factor of $\sim2-3$ for $M_{\star}$ and $\sim3$ or more for SFRs. Sources undetected with \textit{Spitzer}/MIPS and \textit{Herschel}/PACS are indicated explicitly with `IR-undet', to distinguish them from objects in fields without MIPS and PACS observations.\\
$^\dagger$ unreliable IR flux due to blending; see Appendix~\ref{App:IR}.}
\end{deluxetable*}
\capstarttrue

The 10 targets of this study (Table~\ref{tbl:sample}, Figure~\ref{fig:Maps} and \ref{fig:Sample}) are drawn from our SINS/zC-SINF AO program that has yielded AO-SINFONI IFU spectroscopy of the H$\alpha$ and [\ion{N}{2}] emission lines spatially resolved on $\sim1$ kpc scales for 35 massive SFMS galaxies at $z\sim2$ (\citealt{genzel14a, genzel14b, forster-schreiber14, newman14, tacchella15, tacchella15_sci, forster-schreiber18}). The sample is virtually unique, given the long integration times of usually $>10$h to obtain H$\alpha$ spectroscopy at 8 m AO resolution. Our 10 targets were initially taken from various spectroscopic surveys, namely, seven targets are from the zCOSMOS-DEEP survey \citep{lilly07, lilly09}, two targets are from the `BX/BM' sample of \citet{steidel04}, and one target is from the `Deep-3a' survey \citep{kong06}. The specific selection criteria for the SINFONI AO observations were an uncontaminated H$\alpha$ emission line and a minimum expected H$\alpha$ line flux (corresponding roughly to a minimum SFR of $\sim10~M_{\odot}~\mathrm{yr}^{-1}$; \citealt{forster-schreiber09, mancini11, forster-schreiber18}).

In addition to our AO-SINFONI H$\alpha$ data, these galaxies have a wealth of ancillary ground- and space-based multiwavelength data that give integrated stellar masses, global UV+IR SFRs, and other key galactic properties. Furthermore, our $HST$ WFC3 Cycle 19 \# GO12578 program \citep[8 targets;][]{tacchella15} together with our NICMOS pilot study \citep[2 targets;][]{forster-schreiber11a, forster-schreiber11b} has provided rest-frame optical F110W ($J$) and F160W ($H$) data for all targets of this study. These data have given us their rest-frame optical morphology and their $1~\mathrm{kpc}$ distribution of the oldest stellar populations through mass-to-light ratio estimates from the 4000$\mathrm{\AA}$ break \citep[see][]{tacchella15_sci}. Figure~\ref{fig:Maps} presents the data used in this paper. In particular, $HST$ images, dust attenuation maps, UV and H$\alpha$ SFR maps, stellar mass maps, and H$\alpha$ velocity maps are shown. 

Table~\ref{tbl:sample} lists the H$\alpha$ redshift and the main stellar properties. The SED modeling has been presented in \citet{forster-schreiber09, forster-schreiber11a, forster-schreiber11b} and \citet{mancini11}. Briefly, we adopt the best-fit results obtained with the \citet{bruzual03} code, a \citet{chabrier03} initial mass function (IMF), solar metallicity\footnote{We refer to Z=0.02 with solar metallicity throughout this paper, although more recent measurements indicate that solar metallicity may be closer to Z=0.015 \citep{caffau11}.}, the \citet{calzetti00} reddening law, and constant SFRs. We define the stellar mass to be the integral of the past SFR. There are two motivations for doing this: ($i$) this stellar mass remains constant after the galaxy ceases its star formation, which makes the comparison with quiescent galaxies across different epochs simpler; and ($ii$) the sSFR defined with this stellar mass definition is a good indicator for the inverse of the $e$-folding timescale (i.e., roughly the mass-doubling timescale). These are about 0.2 dex larger than the commonly used definition, which subtracts the mass returned to the interstellar medium, i.e., the mass of surviving stars plus compact stellar remnants \citep{carollo13a}. For our sample the stellar masses adopted here (and also in \citealt{tacchella15, tacchella15_sci}) are on average $0.12\pm0.3$ dex higher (with a maximum difference of 0.19 dex) than the ones presented in \citet{forster-schreiber09, forster-schreiber11a, forster-schreiber11b} and \citet{mancini11}. Importantly, all estimates from the literature have been converted to this stellar mass definition. The uncertainty on the stellar mass is a factor of $\sim2-3$, while on the SFRs and stellar ages it is even larger. These uncertainties mainly arise from the basic assumption of the SED modeling, namely, the IMF and the star-formation histories (SFHs). Besides the SED-derived quantities and the UV+IR SFRs, Table~\ref{tbl:sample} presents the integrated values of the 2D maps of dust attenuation and UV SFR (see Section~\ref{sec:measurement} for details).

Our 10 galaxies have stellar masses between $10^{10}~M_{\odot}$ and a few times $10^{11}~M_{\odot}$ and SFRs between $60~M_{\odot}~\mathrm{yr}^{-1}$ and $\sim560~M_{\odot}~\mathrm{yr}^{-1}$. In Figure~\ref{fig:Sample} we show our sample of 10 galaxies in the wider context of the general population of SFGs at the same redshifts (converted to the same stellar mass definition as used in this work). Since our analysis is limited to 10 galaxies, we inevitably probe a limited parameter space of the massive galaxy population at $z\sim2$. We use the SFGs at $z=2.0-2.5$ from the 3D-HST survey \citep{brammer12, skelton14} as our reference sample. Panels $(a)$ and $(b)$ show that our sample probes the typical SFGs on the SFMS at $z\sim2.2$: it lies slightly above the SFMS ridgeline by $0.14^{+0.32}_{-0.30}$ dex. We use the SFMS ridgeline of \citet{whitaker14}, which is based on the $z=2.0-2.5$ star-forming galaxy population from the CANDELS/3D-HST survey \citep{brammer12, skelton14}. In panel $(c)$ we compare the $(U-V)_{\rm rest}$ -- $(U-J)_{\rm rest}$ colors of our sample to the ones of the overall galaxy population drawn from the CANDELS/3D-HST survey at $z=2.0-2.5$ and with $M_{\star}=10^{10.0}-10^{11.5}~M_{\odot}$. We have color-coded the points according to their stellar mass in order to highlight that more massive galaxies are on average redder and more dusty. Our galaxies overlap with the bulk of the SFGs in terms of $(U-V)_{\rm rest}$ -- $(U-J)_{\rm rest}$ colors, but we do not probe massive and dusty SFGs in the upper right corner of the $UVJ$ diagram, which may partially be due to small number statistics in addition to our sample selection criteria. Similarly, looking at the ratio of IR and UV SFR (panel $(d)$), while our sample overlaps with global population traced by the stacking analysis of \citet{whitaker14} of the CANDELS/3D-HST data, our most massive galaxies lie slightly below the ridgeline of SFR$_{\rm IR}$/SFR$_{\rm UV}$ versus $M_{\star}$ of the larger sample. Finally, our sample probes the typical trends in the planes of size (circularized $H$-band half-light radius) versus $M_{\star}$ and S\'{e}rsic index versus $M_{\star}$ for star-forming galaxies at $z=2.0-2.5$ \citep{van-der-wel14a}, as shown in panels $(e)$ and $(f)$, respectively. The gray circles indicate our galaxies with unreliable measurements because the light distributions are not centrally concentrated (assumed S\'{e}rsic models do not represent the galaxy well). See also \citet{tacchella15} for an extended discussion and detailed description of the structural measurements.

Based on \citet[][see also stamps in Figure~\ref{fig:Maps}]{tacchella15}, 9 out of our 10 galaxies are disk galaxies and only one is a clear merger. D3A-15504, Q2343-BX610, ZC405226, and ZC405501 are classified as regular disks, since at the rest-frame optical wavelengths the galaxies show a relatively symmetric morphology featuring a single, isolated peak light distribution and no evidence for multiple luminous components. The velocity maps show clear rotation, and the dispersion maps are centrally peaked; the kinematic maps are fitted well by a disk model with $v_{\rm rot}/\sigma_0\ga1.5$, where $v_{\rm rot}$ is the inclination-corrected rotational velocity and $\sigma_0$ is the velocity dispersion corrected for instrumental broadening and beam smearing. Furthermore, Q2346-BX482, ZC400528, ZC400569, ZC406690, and ZC407302 are classified as irregular disks, because in the optical light the galaxies have two or more distinct peaked sources of comparable magnitude. Their velocity maps show clear signs of rotation. The dispersion maps show a peak, which is, however, shifted in location relative to the centers of the velocity maps. Finally, ZC412369 is classified as a merger, since in the rest-frame optical light two or more distinct peaked sources of comparable magnitude are detected at a projected distance of $\sim5$ kpc from each other. The velocity maps are highly irregular with no evidence for ordered rotation (i.e. $v_{\rm rot}/\sigma_0\la1.5$); the velocity dispersion maps show multiple peaks. 

The active galactic nucleus (AGN) activity in our sample has been discussed in \citet{forster-schreiber14}. None of the sources in our sample are detected in the deepest X-ray observations with Chandra \citep{civano16}; their flux upper limits imply $\log(L_X/\mathrm{erg}~\mathrm{s}^{-1})<42.5$. For some galaxies, an AGN contribution can be argued based on the mid-IR colors (Q2343-BX610), emission lines (D3A-15504), or radio data (ZC400528).

\subsection{New FUV and NUV Data}\label{subsec:Observations}

In our Cycle 22 $HST$ program \# GO13669, we followed up these galaxies with WFC3/UVIS and ACS/WFC observations between December 2014 and June 2015 using the F438W ($B$) and F814W ($I$) filters. Our galaxies lie in a narrow redshift range $2.03<z<2.39$, which puts the FUV into the $B$-band and the NUV into the $I$-band. Therefore, these two passbands cover the entire rest-frame $\sim1200-2700~\mathrm{\AA}$ wavelength window. This gives us the possibility to measure the UV continuum slope $\beta$ in order to constrain the attenuation distribution, one of the main scientific goals of this paper. The WFC3 F438W filter is perfect for the FUV image: the cut-off wavelength of $4000~\mathrm{\AA}$ corresponds to $1230-1270~\mathrm{\AA}$ in the rest frame, the long-wavelength cutoff to $1450-1500~\mathrm{\AA}$. Furthermore, the ratio of H$\alpha$ to rest-frame $1400~\mathrm{\AA}$ (probed with F438W) versus F438W-F160W will enable us to best disentangle extinction and evolutionary effects among clumps and between clumps and interclump regions, which will be addressed in a future publication.

Each target was observed for four orbits with F438W and, if necessary\footnote{We capitalize on the existing COSMOS F814W images for the zCOSMOS targets in the current sample.}, one orbit in F814W, with each orbit split into two exposures with a subpixel dither pattern to ensure good sampling of the PSF and minimize the impact of hot/cold bad pixels and other such artifacts (e.g., cosmic rays, satellite trails). For F438W, the individual exposure time was 1376 s, giving a total on-source integration of 11,008 s, and for F814W, the individual exposure time was 544 s, giving a total on-source integration of 2176 s. 

\subsection{Data Reduction}\label{subsec:DataReduction}

\subsubsection{Charge Transfer Efficiency}

The charge transfer efficiency (CTE) of the WFC3/UVIS detector has significantly degraded over time, as radiation causes permanent damage of the charge transfer device (CCD) detectors. This damage degrades the ability of electrons to transfer from one pixel to another, temporarily trapping electrons during the readout. When uncorrected, the electrons are smeared out in the readout direction, appearing as trails in the images. This affects the photometry and measured morphology of the objects in the images \citep{massey10, rhodes10}. CTE degradation is most severe for low background imaging of faint sources, such as NUV imaging and calibration dark frames, where faint sources or hot pixels can be lost completely. 

As mentioned before, we have one galaxy per pointing, i.e. we were able to choose freely where to place the galaxy on the CCD detector. To reduce the CTE degradation effect on our galaxies, we placed the galaxies close to the readout edge of the CCD. In addition, we apply a pixel-based CTE correction\footnote{\url{http://www.stsci.edu/hst/wfc3/tools/cte_tools}} to the raw data based on empirical modeling of hot pixels in dark exposures \citep{anderson10, massey10}. This correction not only corrects the photometry but also restores the morphology of sources.

\subsubsection{WFC3/UVIS Dark Calibrations}

Dark calibrations are especially important for NUV data because the dark current level in each exposure is high relative to the low sky background. In addition, regular calibration dark data can be used to identify hot pixels, which vary significantly over time. \citet{teplitz13} show that the darks currently provided by STScI are insufficient for data with low background levels after the CTE degradation of WFC3/UVIS. In this paper, we follow the approach of \citet{rafelski15} to improve the dark calibrations.

While the STScI superdarks were mostly sufficient for early data obtained soon after the installation of WFC3, subsequent changes in the characteristics of the detector (such as CTE degradation) increasingly affected the science data. There are three major areas in which the STScI processed superdarks are insufficient, namely, they fail to account for ($i$) all hot pixels, ($ii$) background gradients, and ($iii$) blotchy background patterns. The dark processing methodology used in this paper is presented in detail in the Appendix of \citet{rafelski15}. 

We have used the standard procedure to convert the CTE-corrected raw data to a set of final, flat-fielded, flux-calibrated images (flt-files). We used the Pyraf/STSDAS task \texttt{calwf3} to construct the bad pixel array (data quality array) and to do the bias and dark current subtraction with the new darks. In this step, we have not applied the cosmic-ray rejection since we reject the cosmic rays with \texttt{drizzlepac/astrodrizzle} in a later step.

\subsubsection{Astrometric Alignment}

Here we discuss several sources of astrometric uncertainties in the original data, as well as our approaches to mitigating these and aligning all the images to a common reference grid. The observations were all obtained in a noninteger pixel-offset dither pattern, aimed at ensuring that the PSF was adequately sampled in the final images. Each small angle maneuver introduces a slight uncertainty in positioning (of the order of about $1–2$ mas). In addition, during each orbit the spacecraft undergoes thermal expansion and contraction (`breathing') due to changes in solar illumination, which lead to changes in the optical path length to the detectors, hence resulting in slight scale changes from one exposure to the next. Finally, guide-star reacquisition uncertainties can lead to errors in position, as well as small rotation uncertainties, while a full acquisition of a new guide star has astrometric uncertainties of $\sim0.''3-0.''5$.

We make use of the source positions measured in the WFC3/F110W and F125W ($J$-band) of our previous \textit{HST} program (\# GO12578; \citealt{tacchella15}) as our absolute astrometric reference frame. The alignment was accomplished with \texttt{drizzlepac/tweakreg} using catalog matching, which provides measurements of rotations, as well as removing the bulk of the shifts that are present in the data. This technique was presented in \citet{koekemoer11} and already used in \citet{tacchella15}. The resulting overall astrometric accuracy is $\sim2$ mas in the mean shift positions of all the exposures relative to one another, which is the best possible level that is achievable given the sparse number of sources and their faint flux at these wavelengths.

We use \texttt{drizzlepac/astrodrizzle} to detect cosmic rays and to dither the different exposures to one final image. We choose the same pixel scale as our previous WFC3/F110W, F125W, and F160W images, namely, of $0.05\arcsec$ pixel$^{-1}$, to match the pixel scale of the SINFONI/AO data. This pixel scale provides an adequate sampling of the PSF. Finally, we set pixfrac (defines how much the input pixels are reduced in linear size before being mapped onto the output grid) to 0.8, which was found from experimentation to give the best trade-off between gain in resolution and introduction of rms noise in the final images.

\subsection{Point Spread Functions (PSFs)} \label{subsec:PSF}

In order to derive PSF-matched color maps and PSF-corrected profiles as described below, PSFs of the different bandpasses are required. PSFs for each $HST$ camera are created in slightly different fashions, due to varying constraints of the data.

The creation of the WFC3/NIR data PSFs is described in \citet{tacchella15}. Briefly, we have stacked six well-exposed and nonsaturated stars. The stars are registered to their subpixel centers, normalized, and then co-added via a median. The FWHM is $0.\arcsec16$ and $0.\arcsec17$ for the $J$- and $H$-band, respectively.

The ACS/WFC data and WFC3/UVIS PSFs are created with a hybrid PSF in a similar manner to the PSFs created by \citet{rafelski15}. We follow the approach of the hybrid PSF because the wings of the PSFs cannot be recreated from a stack of stars due to the low number of stars. The PSF model is created with the TinyTim package \citep{krist95}, subsampled to align the PSFs, resampled to our pixel scale, distortion corrected, and then combined with the same dither pattern and drizzle parameters as were used for the data reduction. The ACS/WFC and WFC3/UVIS stacks of stars are created from 4 and 10 unsaturated stars, respectively. The stars are registered to their subpixel centers, normalized, and then co-added via a median. The final hybrid PSF is a combination of the two, composed of the stack of stars up to a radius of 20 pixels ($0-1.\arcsec0$), and of the PSF model from 20 to 75 pixels ($1.\arcsec0-3.\arcsec75$). In order to prevent discontinuities in the resultant PSF, the PSF model and the star are added together, weighted by a smooth transition window with a full width of 20 pixels ($1.\arcsec0$). The resulting FWHM is $0.\arcsec11$ (2.16 pixels) and $0.\arcsec11$ (2.17 pixels) for the ACS/WFC F814W and WFC3/UVIS F438W, respectively.

\section{Methodology for Deriving Dust Attenuation and SFR in \texorpdfstring{\textit{\MakeLowercase{z}}}{z}$\sim2$ SFMS galaxies} \label{sec:method}

\subsection{Estimating the Star-Formation Rate} \label{subsec:Derivation_SFR}

\subsubsection{SFR from UV}

The UV continuum ($1250-2800~\mathrm{\AA}$) intensity of a galaxy is sensitive to massive stars ($\ga5~M_{\odot}$), making it a direct tracer of current SFR. By extrapolating the formation rate of massive stars to lower masses, for an assumed form of the IMF, one can estimate the total SFR.

We adopt the conversions of \citet{kennicutt98}, who assumed a \citet{salpeter55} IMF with mass limits of 0.1 and 100 $M_{\odot}$, and stellar population models with solar metal abundance. Furthermore, they also assumed that the SFH is constant for at least the past 100 Myr. We modify their calibration downward by a factor of 0.23 dex to match a \citet{chabrier03} IMF \citep{madau14}:

\begin{equation}
 \frac{\mathrm{SFR}_{\rm UV}}{M_{\odot}~\mathrm{yr}^{-1}} = 0.82\times10^{-28}~\frac{L_{\rm UV, corr}}{\mathrm{erg}~\mathrm{s}^{-1}~\mathrm{Hz}^{-1}},
 \label{eq:SFR_UV}
\end{equation}
\noindent
where $L_{\rm UV, corr}$ is the UV luminosity at $1500~\mathrm{\AA}$ corrected for attenuation, i.e. $L_{\rm UV, corr} = L_{\rm UV, obs} \times 10^{0.4\mathrm{A}_{\rm UV}}$. We derive $L_{\rm UV, obs}$ from the observed magnitude as follows:

\begin{equation}
 L_{\rm UV, obs} = \frac{4\pi D_L^2(z)10^{-0.4(48.6+m_{\rm UV})}}{1+z},
\end{equation}
\noindent
where $D_L(z)$ is the luminosity distance at $z$, and $m_{\rm UV}$ is the observed
magnitude in the $B$-band. We estimate the amount of UV attenuation, $\mathrm{A}_{\rm UV}$, from the UV continuum slope (i.e. if $f_{\lambda}\propto \lambda^{\beta}$), following \citet{meurer99}:

\begin{equation}
\mathrm{A}_{\rm UV} = 4.43 + 1.99\beta.
\end{equation}

We estimate the UV continuum slope $\beta$ from the FUV-NUV color (observed $B-I$ color), as detailed in Section~\ref{subsec:color_dust}.

\subsubsection{SFR from H$\alpha$}

Another widely used diagnostic for measuring the SFR is nebular emission, with H$\alpha$ being the most common because it is the strongest of the Balmer H recombination lines, and it is least affected by underlying Balmer stellar absorption features and by extinction compared to the higher-order Balmer lines at shorter wavelengths.

The conversion factor between ionizing flux and the SFR is usually computed using an evolutionary synthesis model. Only stars with masses $\ga10~M_{\odot}$ and lifetimes $<20$ Myr contribute significantly to the integrated ionizing flux, so the emission lines provide a nearly instantaneous measure of the SFR, independent of the previous SFH. We again adopt the conversions of \citet{kennicutt98} (Salpeter IMF and solar metal abundance), with the assumption that the SFH is constant for at least the past 10 Myr and that Case B recombination at $T_e=10,000$ K applies. We modify their calibration downward by a factor of 0.23 dex to match a \citet{chabrier03} IMF:

\begin{equation}
 \frac{\mathrm{SFR}_{\rm H\alpha}}{M_{\odot}~\mathrm{yr}^{-1}} = 4.7\times10^{-42}~\frac{L_{\rm H\alpha, corr}}{\mathrm{erg}~\mathrm{s}^{-1}}.
 \label{eq:SFR_Ha}
\end{equation}
\noindent
where $L_{\rm H\alpha, corr}$ is the H$\alpha$ luminosity corrected for attenuation, i.e. $L_{\rm H\alpha, corr} = L_{\rm H\alpha, obs} \times 10^{0.4\mathrm{A}_{\rm H\alpha}}$. 

Since H$\alpha$ lies at $6563~\mathrm{\AA}$, it suffers much less dust extinction compared to the UV rest frame of a galaxy spectrum. On the other hand, the Lyman continuum ionizing radiation that gives rise to the H$\alpha$ emission is mainly produced by massive, short-lived stars that are probably deeply embedded in the giant \ion{H}{2} regions, whereas less massive ($<10~M_{\odot}$) stars still contribute to the UV rest-frame stellar continuum on the long terms, shine over timescales 10 times longer, and have time to clear or migrate out of the dense \ion{H}{2} regions. The net outcome of this process is that H$\alpha$ emission probably suffers from an extra attenuation, parameterized by the $f$-factor that relates $\mathrm{E(B-V)}_{\rm star} = f \times \mathrm{E(B-V)}_{\rm gas}$. 

However, the H$\alpha$ luminosity depends also on the actual extinction in the Lyman continuum, as one gets one H$\alpha$ photon for each Lyman continuum photon that does not get absorbed by a dust grain \citep[see, e.g.,][]{boselli09, puglisi16}. Since we are unable to constrain the extinction in the Lyman continuum, we absorb this part of physics in the overall uncertainty of the $f$-factor (see below), which itself is very uncertain since it depends on the actual extinction law (from the optical to the Lyman continuum) and on the geometry of the emitting regions. 

The extra amount of attenuation suffered by nebular emission is still debated, especially at high redshifts. In the local universe, the stellar continuum undergoes roughly half of the reddening suffered by the ionized gas, namely, $f=0.44$ \citep{calzetti00}. Important to note is that \citet{calzetti00} used two different extinction curves for the nebular \citep[$k(\lambda)_{\rm gas}$;][]{cardelli89, fitzpatrick99} and continuum \citep[$k(\lambda)_{\rm star}$;][]{calzetti00}, which have similar shapes but different normalizations ($R_{\rm V}=3.1$ and 4.05, respectively). If not differently noted, we adopt the same extinction curves here. At higher redshifts, \citet{wuyts13} present a polynomial function to derive extra attenuation from the continuum attenuation; hence, the $f$ value may not necessarily be a constant value for all types of galaxies (see also \citealt{price14}). It is also suggested by recent studies that the typical $f$ value may be higher ($f=0.5-1.0$) for high-redshift galaxies \citep[see, e.g.,][]{erb06c, reddy10, kashino13, koyama15, valentino15, kashino17}.

Because we lack of H$\beta$ data for our galaxies, in this paper we constrain the attenuation of the H$\alpha$ emission line, $\mathrm{A_{\rm H\alpha}}$, by converting the dust attenuation in the continuum at V, $\mathrm{A_{\rm V}}$, using 

\begin{equation}
\begin{split} 
 A_{\rm H\alpha, gas} & = \frac{\mathrm{E(B-V)}_{\rm gas}}{\mathrm{E(B-V)}_{\rm star}} \cdot
 \frac{k(\mathrm{H\alpha})_{\rm gas}}{R_{\rm V, star}} \cdot A_{\rm V, star} \\
 & = \frac{1}{f} \cdot \frac{2.36}{4.05} \cdot A_{\rm V, star}, 
 \label{eq:A_Ha_Calzetti}
\end{split} 
\end{equation}

\noindent with $f=\mathrm{E(B-V)}_{\rm star}/\mathrm{E(B-V)}_{\rm gas}$ ($f=0.44$ from \citealt{calzetti00} and $f=0.7$ from, e.g., \citealt{kashino13}). We note that the prescription of \citet{wuyts13} gives very similar results to this fiducial relation using a value of $f=0.44$.

\subsection{Effects on (FUV$-$NUV) color from Stellar Population Variations} \label{subsec:BI_SP}

In this section we briefly mention the effect of varying the stellar population properties on the (FUV$-$NUV) color. A more detailed discussion including figures can be found in Appendix~\ref{App:color_dust}. 

As highlighted in the Introduction, by far the largest potential effect on the UV color of SFGs is the presence of dust. Therefore, reddening of the UV colors provides a good diagnostic of the magnitude of the dust attenuation. However, other stellar population parameters have also some effect on the UV color. Most importantly, variations in the SFH can lead to a significant effect in the $(\mathrm{FUV}-\mathrm{NUV})$ color. For example, for a given total stellar mass, reducing the age of a constantly star-forming population by a factor of 10 (i.e. form 1 Gyr to 100 Myr), makes a color by $\Delta(\mathrm{FUV}-\mathrm{NUV})\approx-0.1~\mathrm{mag}$ bluer. More importantly, increasing the SFR by a factor of 10 from $10~M_{\odot}/\mathrm{yr}$ to $100~M_{\odot}/\mathrm{yr}$ leads to a bluer $(\mathrm{FUV}-\mathrm{NUV})$ color by about 0.3 mag, while a reduction of the SFR from $100~M_{\odot}/\mathrm{yr}$ to $10~M_{\odot}/\mathrm{yr}$ leads to a 0.3 redder $(\mathrm{FUV}-\mathrm{NUV})$ color (see Appendix~\ref{App:color_dust}) for a given metallicity and dust attenuation. Furthermore, lowering the metallicity from $Z=0.02\rightarrow0.004$ results in $\Delta(\mathrm{FUV}-\mathrm{NUV})\approx-0.1~\mathrm{mag}$. Finally, changing the IMF from Chabrier to Salpeter makes the color redder by a negligible amount ($\Delta(\mathrm{FUV}-\mathrm{NUV})=0.01~\mathrm{mag}$).

Summarizing, for a given dust attenuation, the $(\mathrm{FUV}-\mathrm{NUV})$ can vary owing to variations in the SFH, metallicity, and IMF by at most $\sim0.4$ mag. This age-dust degeneracy has important consequences for the derived SFRs. For a galaxy with an almost completely quenched bulge, we would infer a high dust attenuation and hence would overcorrect the central SFR. Therefore, the quoted central SFRs should be considered as upper limits.

\subsection{From (FUV$-$NUV) Color to Dust Attenuation} \label{subsec:color_dust}

We derive the (FUV$-$NUV)$-\mathrm{A_{\rm V}}$ and (FUV$-$NUV)$-\beta$ conversions in Appendix~\ref{App:color_dust}. Briefly, we obtain the conversions by generating a set of model SEDs from \citet{bruzual03}, for six different metallicities (Z=0.0001, 0.0004, 0.004, 0.008, 0.02, and 0.05) and three different SFHs (constant, exponentially rising, and exponentially decreasing). We find a tight correlations between the (FUV$-$NUV) color and $\mathrm{A_{\rm V}}$ and $\beta$, which are given by Equations~\ref{eq:color_AV} and \ref{eq:color_beta}.

\subsection{Important Remarks about the Dust Attenuation Correction} \label{subsec:caution}

Before proceeding, we want to highlight some key assumptions behind our results and discuss systematic uncertainties in turning the (FUV$-$NUV) color to a dust attenuation. First, it is well known that the chemical and physical properties and the geometrical distribution of dust within external galaxies are a major uncertainty in the evaluation of their dust attenuation properties. In the absence of reliable constraints, we are working here within the traditional framework of the uniform screen approximation. This means that all effects of dust geometry, extinction, and scattering from the interstellar medium and star-forming regions are packaged into an attenuation curve (see, e.g., \citealt{penner15} and \citealt{salmon16} for a more detailed discussion).

Clearly, galaxies are entities in which stars, gas, and dust are spatially mixed. Indeed, in the local universe, \citet{liu13} analyzed the dust distribution of M83 at a resolved spatial scale of $\sim6$ pc, finding that a large diversity of absorber/emitter geometric configurations can account for the data, implying a more complex physical structure than the classical foreground dust screen assumption. However, when averaged over scales of $100-200$ pc, the extinction becomes consistent with the dust screen approximation, suggesting that other geometries tend to be restricted to smaller spatial scales. At high $z$, the dust-star geometry of galaxies is still largely unconstrained. Based on a massive $z=1.53$ SFG, \citet{genzel13} argued that the resolved molecular gas surface density and the resolved attenuation are well modeled by a mixed model of dust and star-forming regions (consistent with \citealt{wuyts11}).

Since in the screen approximation all effects of dust geometry are hidden in the attenuation curve, it is important to have a good understanding of attenuation curve. Unfortunately, it is uncertain whether the attenuation curve should be universal, given that the conditions that produce the attenuation curve are complex and the nature and properties of dust grains may vary from one galaxy to another. They depend on the covering factor, dust grain size (which is dependent on metallicity), and line-of-sight geometry; these can therefore change when, for example, galaxies are viewed at different orientations \citep{witt00, chevallard13} or have different stellar population ages \citep{charlot00}.

In the local universe, we know that the attenuation curve is not universal and varies between the SMC and LMC, the Milky Way, and starburst galaxies. At high $z$, the attenuation curve is less well constrained (see, e.g., \citealt{noll09, kriek13, zeimann15, salmon16, forrest16}). \citet{reddy15} suggest that the attenuation curve for $z\sim2$ galaxies is lower by 20\% in the UV with respect to the Calzetti law, which leads to SFRs that are $\sim20\%$ and stellar masses that are 0.16 dex lower. On the other hand, \citet{mclure18} present a stacking analysis of ALMA data that at $z\sim2.5$ are fully consistent with the IRX$-\beta$ relation expected from the Calzetti law (see also \citealt{bourne17} and \citealt{koprowski18} for consistent results). A detailed analysis of different attenuation curves at different positions within galaxies is beyond the scope of this work and of our data. Therefore, we self-consistently assume the \citet{calzetti00} attenuation curve for $k(\lambda)_{\rm star}$ and \citet{fitzpatrick99} for $k(\lambda)_{\rm gas}$ through this paper, but we highlight that this issue is worth investigating in future work. Overall, assuming a 0.5 dex scatter in the IRX$-\beta$ relation, we infer an uncertainty in A$_{\rm UV}$ of $\sim0.6$ mag, and hence a SFR uncertainty of 0.3 dex. This is of similar order to the systematic uncertainty of the conversion factor from UV (H$\alpha$) luminosity to SFR introduced by the assumed IMF, of the order to $0.2-0.3$ dex, or variations of the SFH that amount to $\sim0.1-0.2$ dex (see Section~\ref{subsec:Derivation_SFR}).

Finally, regions of total dust obscurations would not be detected in our UV analysis. Compact obscured star-formation has been seen in other samples whose selection criteria, as mentioned in the introduction of this paper, are, however, clearly very different from ours \citet{tadaki15, tadaki17} and \citet{barro16}. Specifically for our sample, we present below a strong argument against the presence of totally obscured centers, namely, the good agreement between UV and H$\alpha$ integrated SFRs after dust correction using our dust attenuation maps, and IR-based SFRs obtained from archival \textit{Spitzer} and \textit{Herschel} data.

\section{Measurements of Dust Attenuation and SFR in $z\sim2$ SFMS galaxies}\label{sec:measurement}

In this section, we compare integrated values of the attenuation, UV SFR, and H$\alpha$ SFR maps with the values from the aperture UV-IR photometry. Furthermore, we present the measurements of the dust attenuation profiles.

\subsection{Resolved versus Aperture A$_{\rm V}$}

\begin{figure}
\includegraphics[width=\linewidth]{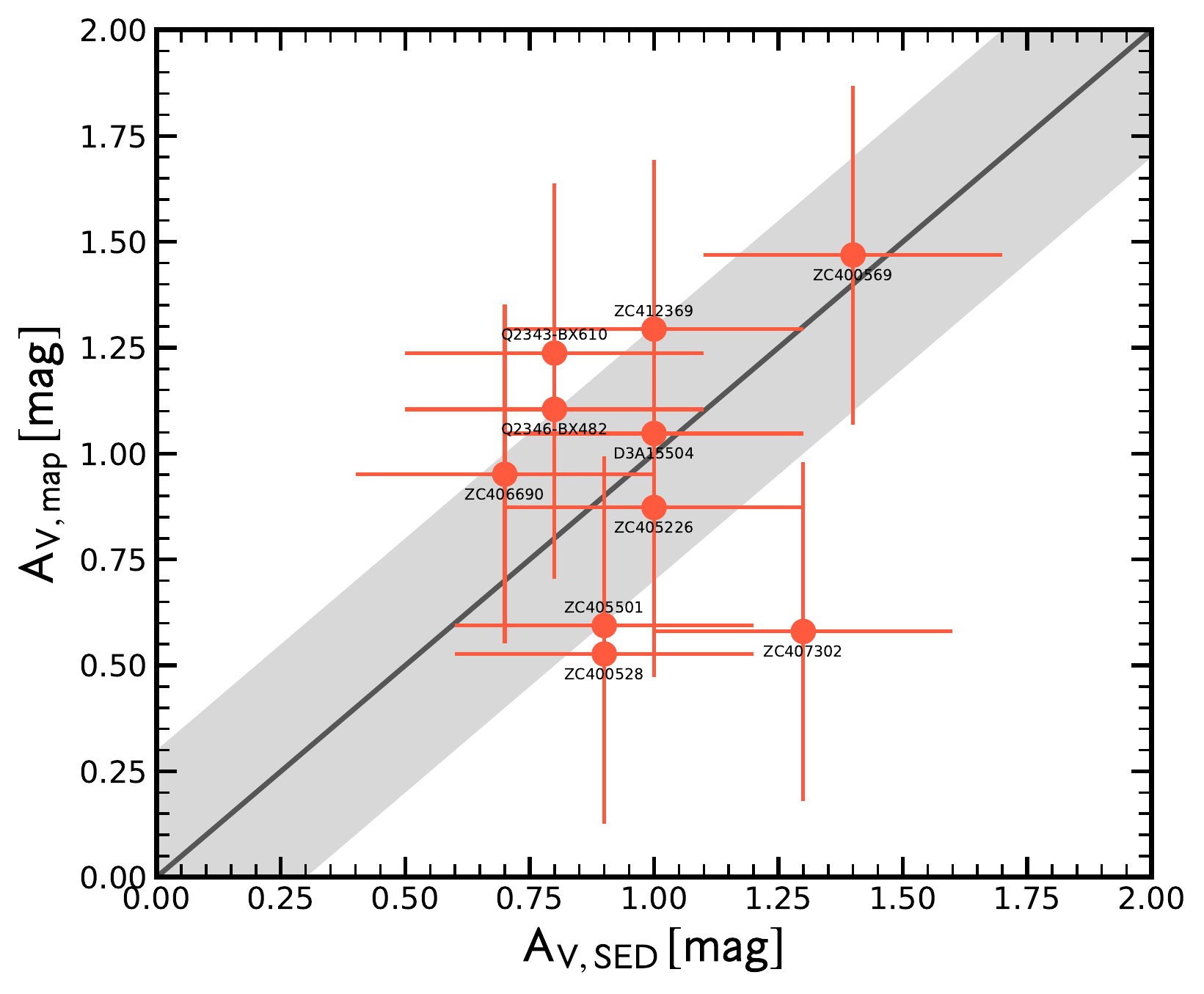} 
\caption{Dust attenuation obtained from the resolved maps (A$_{\rm V, map}$) versus that from the aperture photometry (i.e. SED modeling; $\mathrm{A}_{\rm V, SED}$). The solid black line shows the one-to-one relation. The error bars indicate the systematic errors of $\mathrm{A}_{\rm V, SED}$ and A$_{\rm V, map}$ of 0.3 mag and 0.4 mag, respectively. The attenuation measured by SED fitting and via the integration of the resolved A$_{\rm V, map}$ is broadly similar but correlates poorly on a galaxy-by-galaxy basis (Pearson correlation coefficient of 0.14). The average difference amounts to $-0.01\pm0.35$ mag, i.e. these two estimates are in agreement with each other after considering the relatively large uncertainty.} 
\label{fig:Comparison_AV}
\end{figure}

In Figure~\ref{fig:Comparison_AV} we compare the dust attenuation A$_{\rm V}$ obtained from our maps (A$_{\rm V, map}$) with the one obtained from SED modeling (A$_{\rm V, SED}$), i.e. from the photometry measured in an aperture of 3$\arcsec$. We derive A$_{\rm V, map}$ by summing up all pixels within a 3$\arcsec$ aperture, masking neighboring galaxies, and weighing each pixel according to the flux in the observed \textit{HST} $H$-band image since this band corresponds to the rest-frame $V$-band image.

Overall, the attenuation measured by SED fitting and via the integration of the resolved A$_{\rm V, map}$ is broadly similar but correlates poorly on a galaxy-by-galaxy basis (Pearson correlation coefficient of 0.14). The average difference is $-0.01\pm0.35$ mag, i.e. we find good agreement between A$_{\rm V, map}$ and A$_{\rm V, SED}$, especially after taking into account the relatively large systematic uncertainty of $0.3-0.4$ mag. Seven of our 10 galaxies lie -- within the uncertainty -- close to the one-to-one relation. There are three significant outliers (Q2343-BX610, ZC400528, ZC407302). Both ZC400528 and ZC407302 have close neighbors (see Figure~\ref{fig:Maps}) that are masked in the A$_{\rm V}$ maps but not in the aperture photometry, which can explain the difference. Furthermore, the A$_{\rm V}$ map of Q2343-BX610 shows a large dynamic range, in particular, the center shows a high A$_{\rm V}$ value of $\ga2.5$ mag (see below, Figure~\ref{fig:Dust_Profile_Derivation}). A$_{\rm V}$ from aperture photometry tends to miss these high attenuation values.

\subsection{SFR from IR versus SFR from UV}\label{subsec:SFRIR_vs_SFRUV}

\begin{figure}
\centering
\includegraphics[width=\linewidth]{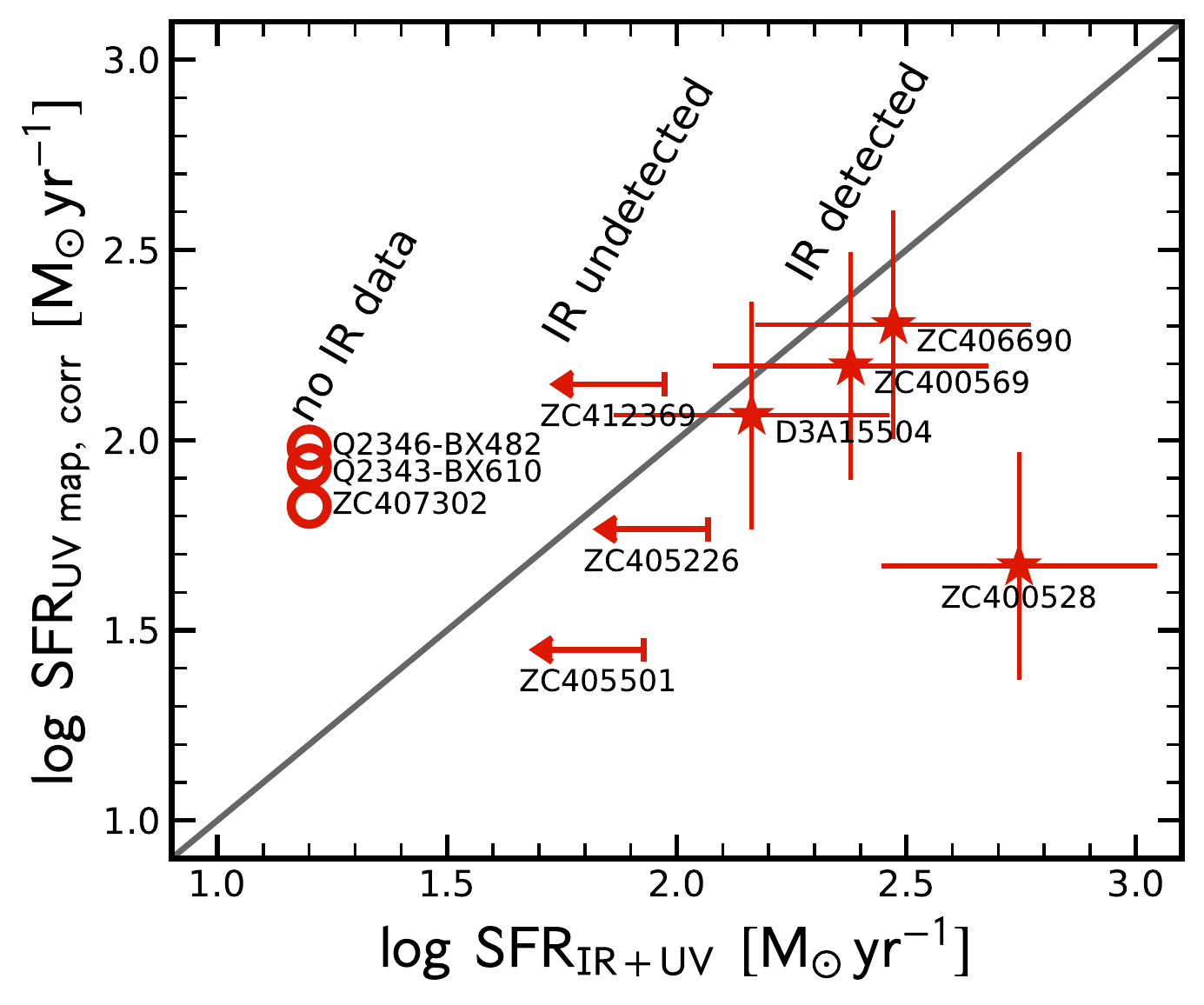}
\caption{Comparison of the SFR estimated from the dust-corrected UV SFR map and the aperture UV+IR photometry. Both SFR estimates agree for three of our four IR-detected objects (D3A15504, ZC400569, ZC406690), while for ZC400528 the UV+IR SFR estimate is significantly higher. For three objects (ZC405226, ZC405501, ZC412369) we have IR upper limits, which are consistent with the SFR estimate from the dust-corrected UV SFR map. For two galaxies (Q2343-BX610 and Q2346-BX482) no IR data is available, while for ZC407302 the IR photometry is not reliable because of a bright neighbor; see Figure~\ref{fig:App_ZC407302}.} 
\label{fig:SFR_UV_vs_IR}
\end{figure}

\begin{figure*}
\includegraphics[width=\textwidth]{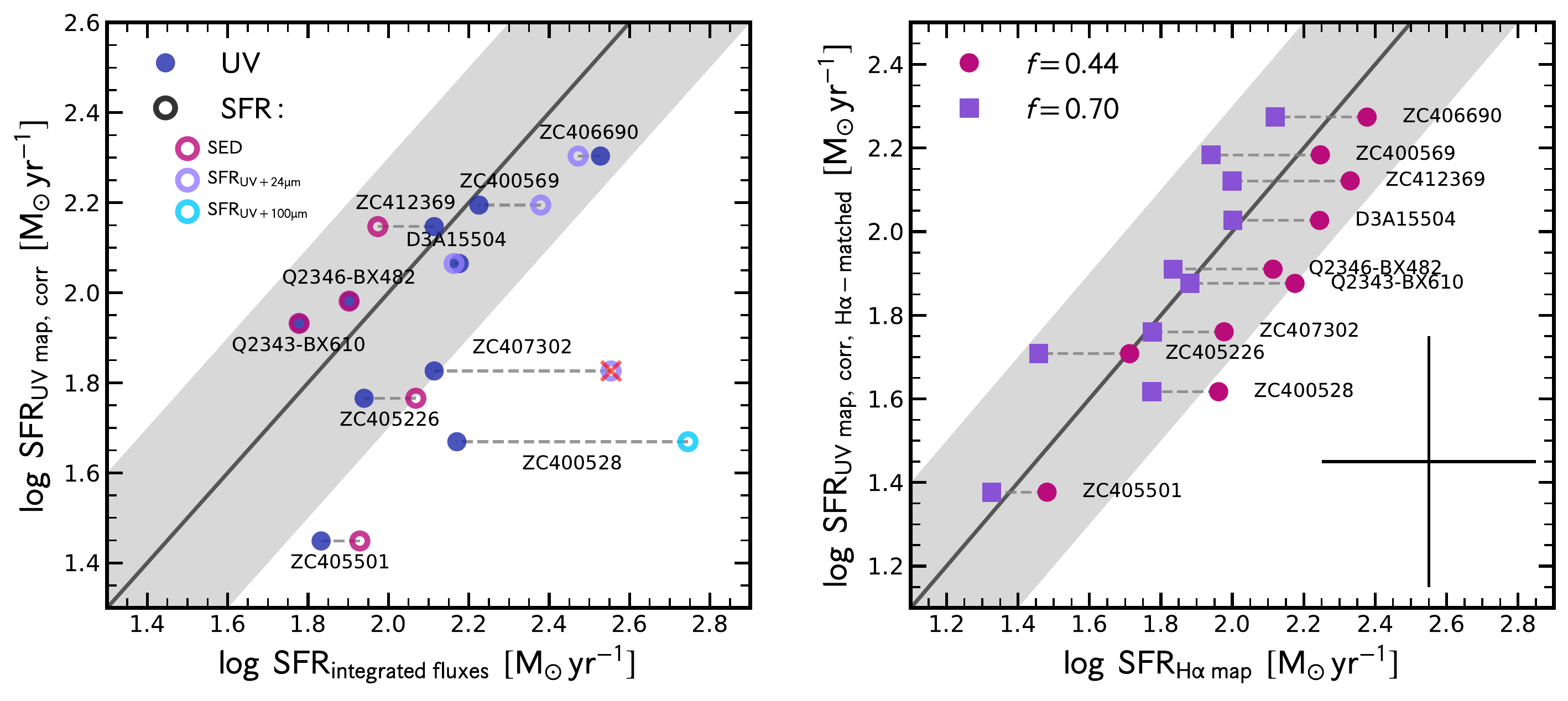}
\caption{Comparison of different SFR estimates. \textit{Left:} SFR estimated from the dust-corrected UV SFR map as a function of SFR estimated from aperture fluxes (UV+IR if available and reliable; otherwise SED), measured on the same 3\arcsec aperture. The black solid line indicates the one-to-one relation and the gray shaded area shows the typical 0.3 dex uncertainty. For most galaxies the SFR estimate from the UV SFR map agrees well with the estimate from the aperture fluxes. Exceptions are ZC400528 and ZC407302: both galaxies have 3-5 times higher SFR predicted from the UV+IR photometry than from UV SFR map. In the case of ZC407302, this can be explained by a neighboring galaxy that contributes significantly to the IR flux. \textit{Right:} SFR estimated from the UV-SFR map as a function of SFR estimated from the H$\alpha$ SFR map, measured on the same H$\alpha$-based aperture. For each galaxy, we show two different H$\alpha$ SFR estimates based on two different attenuation prescriptions: purple circles and violet squares show the $f=0.44$ and $f=0.7$ estimates, respectively. The black solid lines indicate the one-to-one relation. The UV  and H$\alpha$ SFRs agree well within the uncertainties for both attenuation prescriptions, with better agreement for the $f=0.7$ estimates. The error bar on the right indicates the systematic on the inferred SFRs (taking into account variations in the conversion factor of the IMF).} 
\label{fig:Comparison_SFR}
\end{figure*}

The two main SFR indicators used in this paper, the H$\alpha$ recombination line and the UV emission, must be corrected for dust attenuation to recover the intrinsic SFR. In this subsection we use the SFR from IR in order to constrain the intrinsic SFR -- and therefore the effect of dust attenuation -- with an independent method. The IR is a good SFR indicator since, in the limit of high obscuration, most of the UV-optical light from young stars is absorbed and re-emitted into the IR. However, as for other SFR indicators, several assumptions must be made to convert the IR luminosity into an SFR \citep{kennicutt98}. An important caveat of the IR SFR indicator is that dust can also be heated by older stars and/or AGNs. If such dust heating is non-negligible, converting the IR luminosity into an SFR using a standard calibration will overestimate the true SFR \citep[e.g.,][]{sauvage92, smith94, kennicutt09, salim09, kelson10, leroy12, utomo14, hayward14, leja18}. 

Since we do not have spatially resolved IR data, we have to compare our SFR estimates with the ones from IR measured over the whole galaxy. Figure~\ref{fig:SFR_UV_vs_IR} shows a direct comparison of the integrated dust-corrected UV SFR obtained from our resolved maps (SFR$_{\rm UV~map,~corr}$) and total UV+IR SFR, including the upper limits. Furthermore, in Figure~\ref{fig:Comparison_SFR} (left panel), we compare SFR$_{\rm UV~map,~corr}$ both with the  UV SFR (SFR$_{\rm UV}$) and with the total UV+IR SFR (SFR$_{\rm UV+IR}$) obtained from aperture photometry. The SFR$_{\rm UV}$ is derived from the UV rest-frame luminosity, according to the relation of \citet{daddi04}, and adjusted to a Chabrier IMF. For the SFR$_{\rm UV+IR}$, we follow the approach of \citet{wuyts11}: we use either UV + PACS for \textit{Herschel} PACS-detected galaxies \citep[PACS Evolutionary Probe PEP program;][]{lutz11}\footnote{Sources have been extracted using deep \textit{Spitzer} MIPS 24 $\mu$m source positions as priors.} or UV + MIPS 24 $\mu$m for \textit{Spitzer} MIPS-detected galaxies to compute the sum of the obscured and unobscured SFRs. For galaxies lacking an IR detection, the SFR is adopted from the best-fit SED model in Figure~\ref{fig:Comparison_SFR}, while upper limits are plotted in Figure~\ref{fig:SFR_UV_vs_IR}. The values SFR$_{\rm UV~map}$ are obtained by summing up the SFR within the 3\arcsec aperture of the UV SFR map, masking neighboring objects.

Only 1 of our 10 galaxies (ZC400528) is detected at PACS 100 $\mu$m with $\mathrm{SFR}_{\rm UV+100\mu m}=556~M_{\odot}~\mathrm{yr}^{-1}$. This is a factor of $\sim3.8$ and $\sim9$ larger than the dust-corrected UV SFR estimate from the aperture photometry and the integrated map, respectively. As shown in the Appendix~\ref{App:IR}, ZC400528 has a close but faint neighbor galaxy that is not resolved as an individual source in the \textit{Spitzer} MIPS 24 $\mu$m data. Furthermore, this is the only source that is detected with the VLA \citep{schinnerer10} with a 1.4 GHz flux density of 67 $\mu$Jy that would imply an SFR$\approx790~M_{\odot}~\mathrm{yr}^ {-1}$. Together with the broad component in H$\alpha$, [\ion{N}{2}] and [\ion{S}{2}] emission lines, with a typical velocity FWHM of 1500 km/s, and an [\ion{N}{2}]/H$\alpha$ ratio of $\sim0.6$ measured in this galaxy \citep{forster-schreiber14, forster-schreiber18}, this points to AGN activity. However, it is unclear how much an AGN could contribute to the IR and radio measurements. Hence, this measurement of the UV+IR SFR should be interpreted with caution. Mapping of the dust continuum emission with high spatial resolution will provide further insight in the future.

Four galaxies have been detected in \textit{Spitzer} MIPS 24 $\mu$m and we estimated $\mathrm{SFR}_{\rm UV+24\mu m}$. For three galaxies (D3A15504, ZC400569, ZC406690), these estimates agree well with the dust-corrected UV SFR estimates. On average, the UV+IR SFR is $0.15\pm0.04$ dex higher than the SFR estimated from the UV maps, which is well within the uncertainty of 0.3 dex. For one galaxy (ZC407302) the SFR estimated from UV+24 $\mu$m is a factor of three higher than the UV estimate (see Figure~\ref{fig:Comparison_SFR}). This is explained by source confusion: ZC407302 has a close, low-$z$ neighbor that contributes significantly to the 24 $\mu$m flux (see Appendix~\ref{App:IR}). Therefore, we ignore IR measurement for this galaxy and indicate in Figure~\ref{fig:SFR_UV_vs_IR} that this galaxy does not have any reliable IR data. We use the dust-corrected UV SFR estimate as the total SFR instead of the UV+IR SFR for this galaxy throughout this paper.

Moreover, three zCOSMOS galaxies (ZC405226, ZC405501, and ZC412369) have no IR detection and upper limits of $\sim100~M_{\odot}~\mathrm{yr}^{-1}$. For ZC405226 and ZC412369, this indicates within the errors that the SFR hidden in large amounts of dust is negligible. For ZC405501, the IR upper limit is 0.5 dex above the SFR estimated from the UV map, and hence additional star formation cannot be ruled out. Finally, two galaxies (Q2343-BX610 and Q2346-BX482) have no IR coverage. For all galaxies that have been not detected in the IR we can estimate the upper limit on the SFR. The PEP COSMOS field is observed for 200 h to a 3$\sigma$ depth at 160 $\mu$m of 10.2 mJy, reaching at $z\sim2$ IR luminosities of a few times $\sim10^{12}~L_{\odot}$ (SFR about $350~M_{\odot}~\mathrm{yr}^{-1}$). A more stringent upper limit can be obtained for the COSMOS MIPS 24 $\mu$m data: the flux limit of these data is $\sim80$ $\mu$Jy, which corresponds at $z\sim2$ to an SFR of about $150~M_{\odot}~\mathrm{yr}^{-1}$. 

Summing up, 6 of our 10 galaxies have only weak or no detection in the IR, 3 galaxies have no or not reliable IR coverage, and 1 galaxy (ZC400528) has a clearly higher UV+IR SFR than dust-corrected UV SFR. ZC400528 has a faint neighbor and also shows features that are consistent with AGN activity, making the IR flux difficult to interpret. Nevertheless, we use UV+IR SFR as a fiducial SFR estimate for this galaxy. Overall, we conclude that, in our sample, there is no evidence for substantially more star formation hidden by dust relative to what we infer from our dust-corrected, resolved UV and H$\alpha$ SFRs. This is consistent with the overall sample properties shown in Section~\ref{sec:Sample_Data}, where we show that our galaxies are not very dusty. This gives us confidence that correcting UV and H$\alpha$ SFR profiles with the UV-based dust attenuation profiles that we derive in this paper will return reliable estimates of where, spatially resolved within galaxies, most of the star-formation activity is localized.

\subsection{SFR from UV versus SFR from H$\alpha$}\label{subsec:UV_vs_Ha}

We compare the integrated SFR maps from UV and H$\alpha$ in Figure~\ref{fig:Comparison_SFR} (right panel). Since the H$\alpha$ maps do not extend over the full 3$\arcsec$ aperture (see Figure~\ref{fig:Maps}) we have to match the UV and H$\alpha$ apertures and PSFs. We define the H$\alpha$ aperture by the pixels with signal-to noise ratio (S/N) $\geq3$. Clearly, the measured SFR in this aperture is not fully representative of the total SFR, but it is good for the purpose of comparing the UV with the H$\alpha$ SFR map. 

As mentioned in Section~\ref{subsec:color_dust}, correcting the H$\alpha$ SFR estimate for dust attenuation (A$_{\rm H\alpha, gas}$; Equation \ref{eq:A_Ha_Calzetti}) depends on the adopted prescription. In Figure~\ref{fig:Comparison_SFR} we show the SFR$_{\rm H\alpha, map}$ assuming $f=0.44$ and $f=0.70$. Overall, the SFRs estimated from UV and H$\alpha$ agree well. The average log-difference between SFR$_{\rm UV, map}$ and SFR$_{\rm H\alpha, map}$ is $0.21^{+0.05}_{-0.13}$ dex and $-0.06^{+0.07}_{-0.14}$ dex for $f=0.44$ and $f=0.70$, respectively. The SFR$_{\rm H\alpha, map}$ with $f=0.70$ slightly underpredicts the SFR$_{\rm UV, map}$ values, while the SFR$_{\rm H\alpha, map}$ with $f=0.44$ slightly overpredicts the SFR$_{\rm UV, map}$ values. We use $f=0.70$ as our fiducial conversion factor for the rest of the paper.

It is actually unclear whether the UV and the H$\alpha$ SFRs must be the same since they probe different timescales. In particular, as discussed in Section~\ref{subsec:Derivation_SFR}, H$\alpha$ probes mainly $\sim10^7$ yr old stars, while the UV-derived SFR is somewhat sensitive to the SFH over a $\sim10$ times longer interval. Overall, considering the uncertainty from the $f$-factor together with the possibility of probing the SFRs on different timescales, we find that our H$\alpha$ SFR with both $f=0.44$ and $f=0.70$ are in the same ballpark as expectations from the UV SFRs.

\subsection{PSF-corrected Dust Attenuation Radial Profiles}\label{subsec:dust_profiles}

\begin{figure*}
\includegraphics[width=\textwidth]{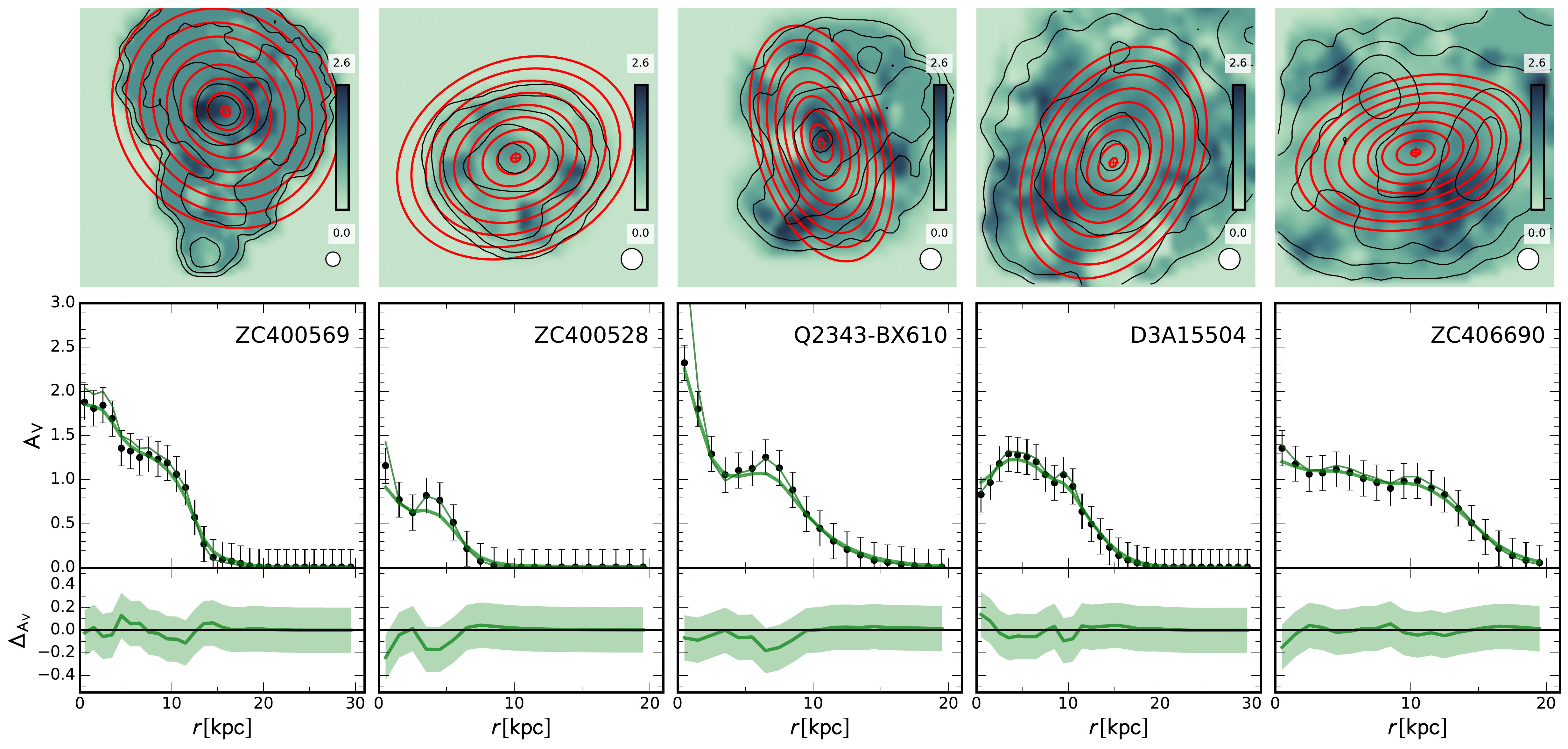}
\hspace{5. mm}
\includegraphics[width=\textwidth]{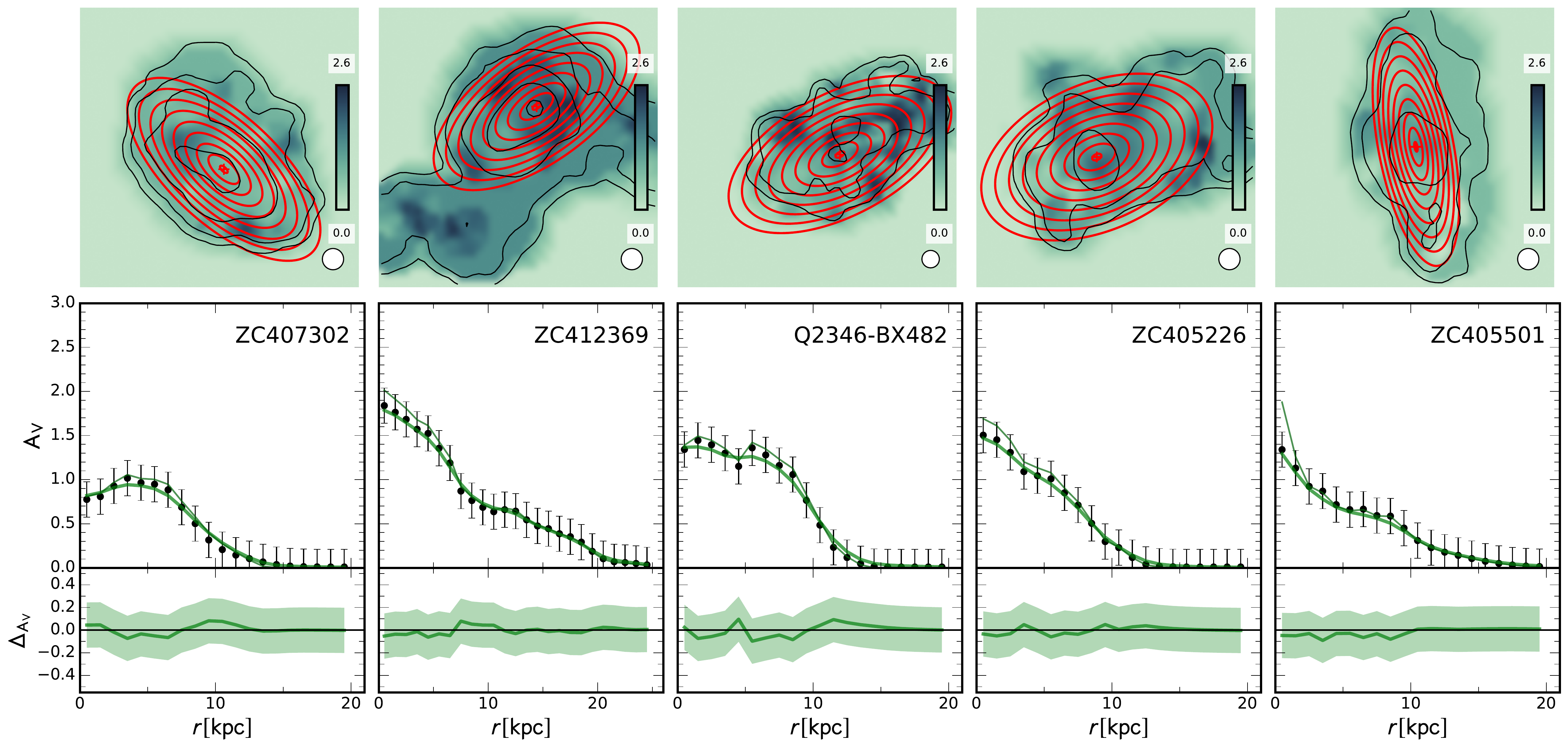}
\caption{Derivation of the azimuthally averaged dust attenuation $\mathrm{A_{\rm V}}$ profiles. For each galaxy, the top panel shows the $\mathrm{A_{\rm V}}$ map. The red ellipses show the apertures used to construct the azimuthally averaged profile and the contours show the mass distribution. The middle panels show the $\mathrm{A_{\rm V}}$ profiles: the data points with error bars indicate the measured values directly from the $\mathrm{A_{\rm V}}$ map, while the thin and thick green lines show the PSF-corrected and PSF-convolved profiles, respectively. The bottom panels show the difference between the data and the PSF-convolved profiles. The PSF-corrected profiles describe the data well with less than 0.2 mag difference, which is within the estimated uncertainty.} 
\label{fig:Dust_Profile_Derivation}
\end{figure*}

To study the variation of extinction within galaxies, we used the rest-frame FUV and NUV (observed $B$- and $I$-band) images to construct a 2D (FUV$-$NUV) color map for each galaxy, following a similar procedure to that used \citet{tacchella15} for estimating the mass surface density distribution from the $J-H$ color map. Since the FUV and NUV observations have different PSFs, we cross-convolved each passband with the PSF of the other passband. To increase the S/N of the color maps in the outer regions of galaxies, where the flux from the sky background is dominant, we performed an adaptive local binning of pixels using a Voronoi tessellation approach, using the publicly available code of \citet{cappellari03}. These color maps are then converted to $\mathrm{A_{\rm V}}$ maps using Equation~\ref{eq:color_AV}, which are shown in Figure~\ref{fig:Dust_Profile_Derivation}. 

In order to compute the PSF-corrected $\mathrm{A_{\rm V}}$ profiles, we measure the 1D profiles in elliptical apertures with the same ellipticity and center as obtained by the rest-frame optical (observed $H$-band) image GALFIT fits \citep{tacchella15}. At each radius, the profile value consists of the median of all $\mathrm{A_{\rm V}}$ at that radius. Taking the mean instead of the median does not change the profiles. Furthermore, using the $B$- and $I$-band light profiles to derive the color and then the $\mathrm{A_{\rm V}}$ profiles also produces similar results. 

These 1D $\mathrm{A_{\rm V}}$ profiles were then fitted with a \citet{sersic68} profile taking into account each galaxy's PSF. We choose to use a S\'{e}rsic function for modeling the $\mathrm{A_{\rm V}}$ profiles simply as a mathematical representation of the data and because the three free parameters in the S\'{e}rsic function give a high flexibility. Although the $\mathrm{A_{\rm V}}$ profiles are shallower than the typical galaxy surface brightness profiles (and also the SFR and stellar mass surface density profiles), they are still well described by a S\'{e}rsic profile. In order to correct the profiles for deviation from the best-fit single S\'{e}rsic model, we derive the residual profile and add this to the best-fit deconvolved S\'{e}rsic model, to derive the residual-corrected profiles (Figure~\ref{fig:Dust_Profile_Derivation}). Overall, the PSF-convolved model reproduces the data well, with differences $\la0.2~\mathrm{mag}$.

A center of a galaxy must be chosen for constructing a radial profile. We have extensively discussed the influence on the profiles of choosing different centers in \citet{tacchella15}, since it sets the foundation for the physical interpretation. In particular, we have highlighted the differences between dynamical, mass-weighted, and light-weighted (rest-frame optical light) centers. For most galaxies, these centers agree with each other (differences are $\la0.4$ kpc, i.e. $\la1$ pixel), and we assume the mass-weighted center as our fiducial center. An exception is galaxy ZC406690, which shows a clumpy morphology and for which the center is ambiguous. The kinematic center lies in the center of a ring-like structure, on which the field of view of SINFONI IFU is centered. The center of mass lies about 5 kpc to the east, on the largest clump visible in Figure~\ref{fig:Maps}. For this galaxy, we assume the dynamical center determined from the H$\alpha$ kinematics in order to be consistent with previous studies \citep[e.g.,][]{genzel14a, genzel14b, forster-schreiber14, newman14, tacchella15, tacchella15_sci, forster-schreiber18} and to be able to compare the UV and H$\alpha$ emission on scales out to $\sim10$ kpc.

\begin{figure*}
\includegraphics[width=\textwidth]{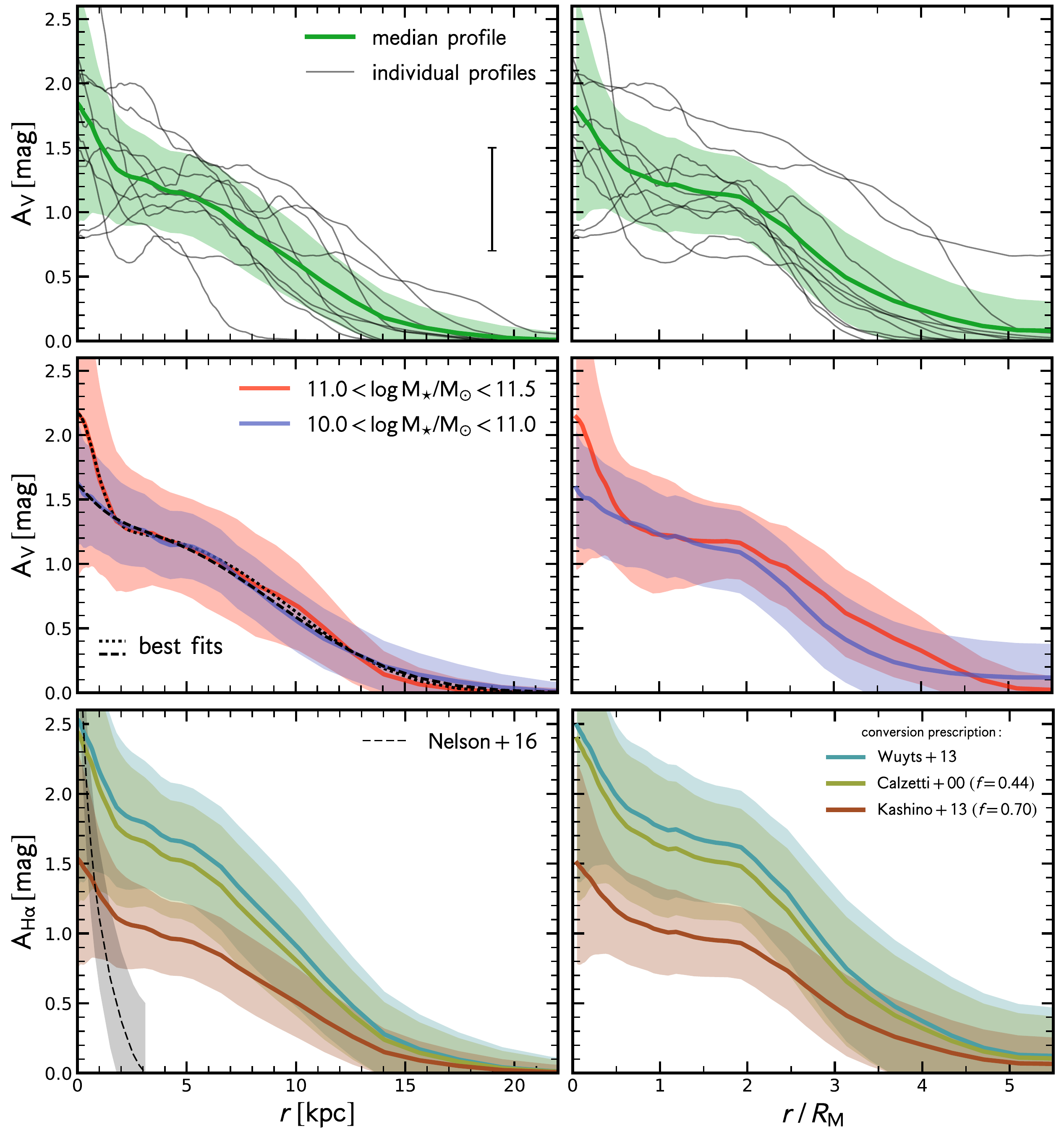} 
\caption{Average dust attenuation profiles. Left panels show the average profiles with radius in physical units of kpc, and right panels show the average profiles in units of radius normalized by the half-mass radius ($R_{\rm M}$). In the top panels, the thin gray lines show the individual $\mathrm{A}_{\rm V}$ profiles, highlighting the substantial variety of the attenuation profile from galaxy to galaxy. The green solid lines indicate the average $\mathrm{A}_{\rm V}$ profiles with their $1\sigma$ scatter. The error bar indicates the 0.4 mag uncertainty stemming from the UV color to A$_{\rm V}$ conversion due to stellar population variations. The middle panels show in cyan and orange the average $\mathrm{A}_{\rm V}$ profiles for the lower- and higher-mass galaxies, respectively. The average $\mathrm{A}_{\rm V}$ profile increases toward to the center. The bottom panels show the inferred dust attenuation toward HII regions ($\mathrm{A}_{\rm H\alpha}$) averaged over all galaxies. The solid lines show the inferred $\mathrm{A}_{\rm H\alpha}$ profiles assuming different conversion prescriptions (see legend). The dashed line indicates the recent measurements from the Balmer decrement for a stack of galaxies with $\log~M_{\star}/M_{\odot}=9.8-11.0$ at $z\sim1.4$ \citep{nelson16_balmer}. } 
\label{fig:Dust_Profile}
\end{figure*}

Figure~\ref{fig:Dust_Profile} shows the individual and stacked $\mathrm{A_{\rm V}}$ profiles as well as $\mathrm{A_{\rm H\alpha}}$ profiles. The top panels of Figure~\ref{fig:Dust_Profile} show substantial galaxy-to-galaxy variations in the attenuation profiles. The variation in attenuation at a given radius (physical or scaled) amounts to about 1.5 mag for different galaxies. Most galaxies show increasing attenuation profiles within the inner 5 kpc toward the centers, while however a few show the opposite trend. These differences reflect different SFR profiles, as most of the star-formation can happen in star-forming clumps in, near or far away from the galaxy centers.

The stacked profile shows that galaxies have on average an attenuation profile that rises toward the center to $\sim1.8~\mathrm{mag}$, with a significant amount of dust attenuation ($\sim0.6~\mathrm{mag}$) out to 10 kpc. \citet{wuyts12} found a similar decline in star-forming galaxies with $M_{\star}>10^{10}~M_{\odot}$ at $0.5<z<1.5$, but no explicit radial profiles were presented. Furthermore, these rather weak radial trends agree with local SFGs, where continuum extinction was found from $\sim1.3~\mathrm{mag}$ to $\sim0.8~\mathrm{mag}$ from the center to the optical radius $R_{25}$ \citep{munoz-mateos09}.

Splitting our sample into two mass bins ($10.0<\log~M_{\star}/M_{\odot}<11.0$ and $11.0<\log~M_{\star}/M_{\odot}<11.5$), we find that there is not a large difference between the two mass bins. As a way of quantifying the slopes and normalizations of the attenuation profiles, we parametrise them with a two-component S\'{e}rsic fits (without implying that this specific profile shape bears any physical meaning for the diagnostic in question). The outer (or `disk') component has an $n\approx0.3-0.4$ and $R_{\rm e}\approx 8~\mathrm{kpc}$ for both mass bins; the inner (or `bulge') components differ instead for the high- and low-mass bins, yielding respectively $n\approx0.5$ and $R_{\rm e}\approx 1.0~\mathrm{kpc}$, and $n\approx0.8$ and $R_{\rm e}\approx 1.7~\mathrm{kpc}$.

Figure~\ref{fig:Dust_Profile}, bottom panels, show the median dust attenuation of the ionized gas at H$\alpha$, $\mathrm{A_{\rm H\alpha}}$. For converting the $\mathrm{A_{\rm V}}$ profiles to $\mathrm{A_{\rm H\alpha}}$ profiles we use the procedure described in Section~\ref{subsec:color_dust} and in particular Equation~\ref{eq:A_Ha_Calzetti}. Note that, by construction, using $f=0.44$ (or, similarly, the prescription of \citealt{wuyts13}) increases of $\sim 0.5~\mathrm{mag}$ the attenuation for nebular line regions at H$\alpha$ relative to the stellar continuum at V; for $f=0.70$ the difference amounts instead to about $-0.2~\mathrm{mag}$. 

We compare our profiles with the results of \citet{nelson16_balmer} at $z\sim1.4$, which are based on a Balmer decrement analysis of the stack of several hundred 3D-HST galaxies with masses in the range of $10^{9.8}-10^{11.0}~M_{\odot}$, a sample that is much more representative than our sample of 10 galaxies, albeit at lower redshift (see also \citealt{nelson16_insideout} for a discussion on the SFR profiles in these galaxies). The key finding of 3D-HST analysis is that $M_{\star}=10^{9.8}-10^{11}~M_{\odot}$ galaxies have high central dust attenuation. In more detail, their average $\mathrm{A_{\rm H\alpha}}$ profile exhibits a sharp peak in the galaxy center but it is nearly transparent beyond 2 kpc. In the central $\sim1$ kpc region, the $\mathrm{A_{\rm H\alpha}}$ estimates roughly agree, while the radial trend from the Balmer decrement analysis is much steeper than our slowly declining profiles. The Balmer decrement is measured at longer wavelengths than our estimate that comes from the UV and thus it probes higher optical depths. We measure similar attenuation values in the central kiloparsec region. The differences of the profiles at $2-3$ kpc are difficult to interpret, also because the uncertainties are larger. One could argue for a gradient in the additional dust attenuation toward the star-forming region (i.e. $f=0.44$ in the central part, while $f=1.0$ in the outskirts). Following the discussion by \citet{wang17_dust}, this difference in inferred $\mathrm{A_{\rm H\alpha}}$ profile could reflect blending the [\ion{N}{2}] and H$\alpha$ in the \textit{HST} grism data, in appropriate SFH assumptions, and different stacking methods (f.e. \citealt{nelson16_balmer} is normalizing by F140W flux; hence, high equivalent width regions are more highly weighted, leading to a decrease the inferred attenuation). \citet{nelson16_balmer} considered [\ion{N}{2}] contamination but concluded that it does not affect their results even if there was an [\ion{N}{2}]/H$\alpha$ gradient as steep as in local SFGs. Overall, it is unlikely that the full difference can be explained by such tweaks. Spatially resolved maps of the Balmer decrement and the UV color in combination with submillimeter dust maps of individual galaxies will shed more light on this in the future.

\subsection{Radial Star-Formation Rate Density Profiles}\label{subsec:SFR_Profiles}

\begin{figure*}
\includegraphics[width=\textwidth]{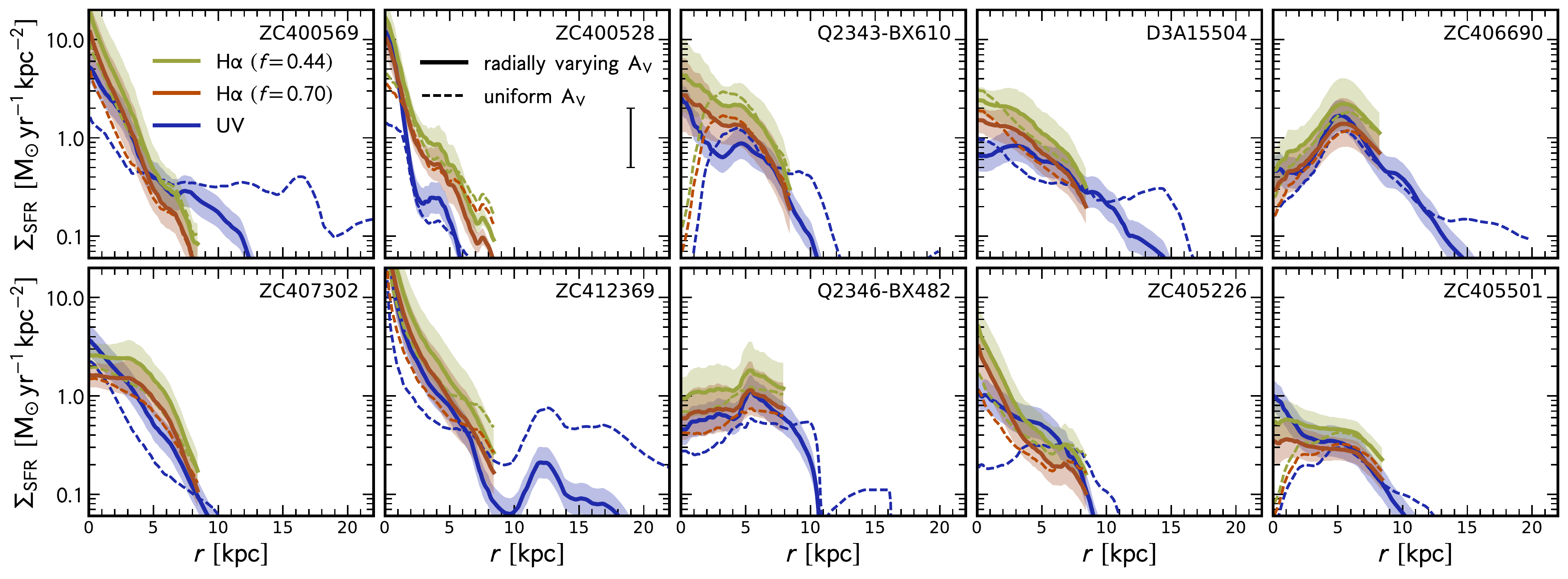} 
\caption{Radial SFR surface density profiles. Each panel shows the profiles for an individual galaxy. The khaki, red, and blue lines indicate the SFR estimated from H$\alpha$ with $f=0.44$, H$\alpha$ with $f=0.70$, and UV, respectively. The dashed and solid lines show the SFR profiles assuming a uniform dust screen and the radial $\mathrm{A_{\rm V}}$ profiles from Figure~\ref{fig:Dust_Profile}, respectively. The UV and H$\alpha$ (especially with $f=0.70$) SFR profiles agree within the uncertainty. The error bar in the second panel indicates the typical systematic error on the inferred SFR ($\approx0.3$ dex, i.e. including IMF variations). Our galaxies have SFR densities in their centers of $0.3-20~M_{\odot}~\mathrm{yr}^{-1}~\mathrm{kpc}^{-2}$. Several galaxies have SFR density peaks in their outskirts that arise from heavily star-forming clumps. Since most of the attenuation profiles are rising toward the centers, the SFR profiles using the radial-dependent $\mathrm{A_{\rm V}}$ correction are more centrally concentrated than the SFR using the uniform $\mathrm{A_{\rm V}}$ correction. }
\label{fig:SFR_Profiles}
\end{figure*}

With the dust attenuation maps and profiles in hand, we are now able to correct the radial SFR density profiles, improving on the analysis of \citet{tacchella15_sci} where a uniform dust attenuation correction throughout the galaxy was applied to the H$\alpha$ flux. In this section, we present the SFR profiles and how they are affected by different dust attenuation corrections. 

In Figure~\ref{fig:SFR_Profiles} we present the radial H$\alpha$ and UV SFR surface density profiles. We derive the radial SFR surface density profiles in the same way as the aforementioned dust attenuation profiles (see Section~\ref{subsec:dust_profiles}). 

Overall, there is a large diversity in the SFR profile shapes. Although on average SFR surface density profiles are well represented by an exponential star-forming disk of different sizes, there is richness of behavior in the individual profiles. Some galaxies have steeper profiles (ZC400528 and ZC412369) or flatter profiles due to bumpy features in the outskirts. The bumpy features can be explained by star-forming clumps in the galaxies' outskirts. Two exceptional galaxies are Q2346-BX482 and ZC406690, which show nearly flat or even increasing SFR density profiles toward the outskirts. Galaxy Q2346-BX482 has several star-forming clumps at similar galactocentric distances that together surpass the SFR density in the center (see Figure~\ref{fig:Maps}). Galaxy ZC406690 shows a reduction in the SFR density in the center. The SFR density peaks at $\sim5$ kpc. As mentioned before, this signature must be interpreted with care. We adopted as our fiducial center the kinematic center. Using instead the mass-weighted center (which also coincides with the peak of SFR) would make the profile centrally peaked, i.e. there would be no reduction in SFR toward the center.

We measure a typical inner ($<1~\mathrm{kpc}$) SFR density of $\Sigma_{\rm SFR, 1kpc}\approx2-12~M_{\odot}~\mathrm{yr}^{-1}~\mathrm{kpc}^{-2}$ for the more massive galaxies in our sample ($11.0<\log~M_{\star}/M_{\odot}<11.5$; namely, ZC400569, ZC400528, Q2343-BX610, and D3A15504), while the lower-mass galaxies ($10.0<\log~M_{\star}/M_{\odot}<11.0$) have $\Sigma_{\rm SFR, 1kpc}\approx0.3-3~M_{\odot}~\mathrm{yr}^{-1}~\mathrm{kpc}^{-2}$. In the lower-mass bin, we have excluded ZC412369 with $M_{\star}\approx3\times10^{10}~M_{\odot}$ which is a merger, causing probably the enhanced SFR in the center of $\Sigma_{\rm SFR, 1kpc}\approx20~M_{\odot}~\mathrm{yr}^{-1}~\mathrm{kpc}^{-2}$. These quoted $\Sigma_{\rm SFR, 1kpc}$ can be considered as upper limits (in particular of the most massive galaxies in our sample) because of the age-dust degeneracy that would lead us to infer high attenuation values for galaxies with prominent, nearly quiescent bulges.

Focusing now on the difference between H$\alpha$ and UV SFR density profiles, we find overall good agreement between the estimators. As highlighted in Section~\ref{subsec:UV_vs_Ha}, the difference in the overall normalization between H$\alpha$ and UV SFRs depends on the $f$-factor. The H$\alpha$ and UV SFRs agree more with $f=0.70$ than with $f=0.44$. 

Since the overall normalization difference between H$\alpha$ and UV is mainly determined by the dust prescription, it may be more compelling to analyze the difference in the profile shapes. For most galaxies, there are small differences in the profile shape. Some galaxies show steeper SFR profiles in H$\alpha$ than in UV (D3A15504, ZC400569, ZC405226), while others show the contrary (Q2343-BX610, ZC400528, ZC405501, ZC407302). More concentrated UV profiles may indicate the progression of the star formation toward to the outskirts.

Finally, we can also quantify the difference between using the radially varying A$_{\rm V}$ profiles and using a uniform A$_{\rm V}$ value (fiducial assumption in the work of \citealt{tacchella15_sci}) for correcting the SFR density for dust attenuation. In Figure~\ref{fig:SFR_Profiles} these two different dust corrections are shown with solid and dashed lines, respectively. Since the A$_{\rm V}$ profiles are increasing toward the centers, it is understandable that correcting the SFR density by those profiles leads to more centrally concentrated SFR density profiles than when using a uniform A$_{\rm V}$ value. In two cases (Q2343-BX610 and ZC405501), the A$_{\rm V}$ profiles are so steep that the SFR density profiles reverse, i.e. they are increasing toward the center with the radially varying A$_{\rm V}$ profiles, while they are decreasing with the uniform A$_{\rm V}$ value.

\section{Discussion}\label{sec:discussion}

We discuss in this section some implications for relative disk versus bulge growth. In particular, we interpret our empirical results specifically in the framework of our previous observational and numerical work.

\subsection{The making of massive spheroids on the $z\sim2$ star-forming Main Sequence}\label{subsec:sSFR_Profiles}

\begin{figure*}
\includegraphics[width=\textwidth]{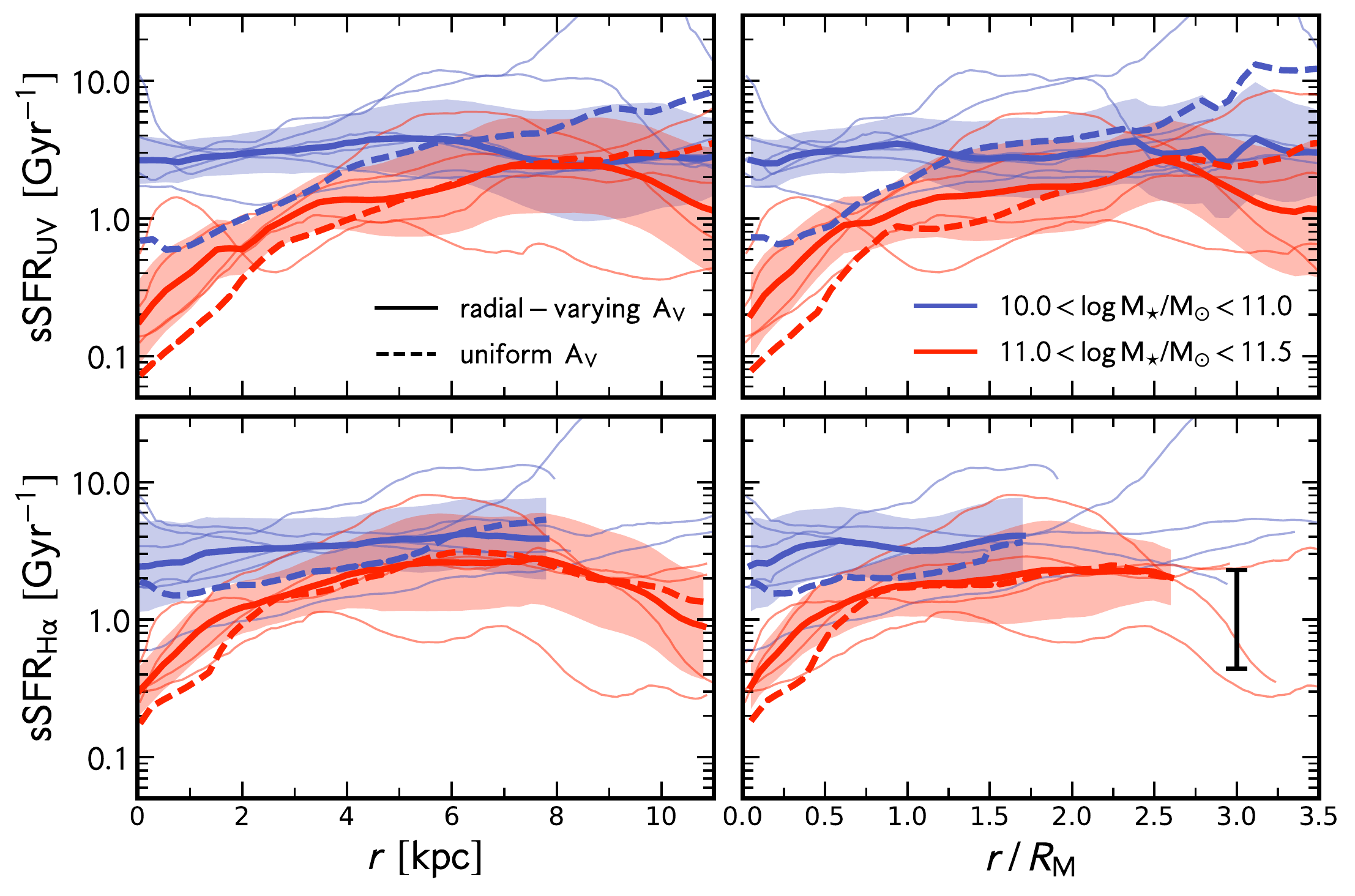} 
\caption{The sSFR profiles from UV (top) and H$\alpha$ (assuming $f=0.70$; bottom) are shown as a function of radius in physical units (left) and scaled by the half-mass radius $R_{\rm M}$ (right). The color coding shows the two mass bins. The error bar in the bottom right panel indicates the typical systematic error on the inferred sSFR at $\mathrm{sSFR}=1~\mathrm{Gyr}^{-1}$ ($\approx0.4$ dex, i. e. including variations of the IMF). The thin lines show the individual galaxies, while the solid and dashed lines show the average profiles dust-corrected respectively using the radially-varying attenuation profiles and the uniform dust screen model. The radially-varying attenuation correction increases the sSFR in the central regions, but there is only little change in the overall sSFR profiles' shapes. Nevertheless, in massive galaxies these profiles remain centrally-suppressed relative to the galaxies outskirts.}
\label{fig:sSFR_profiles}
\end{figure*}

The SFR profiles presented in Section~\ref{subsec:SFR_Profiles} inform us on the location where $z\sim2$ SFMS galaxies sustain their high SFRs. However, the newly formed stars might be, in principle, a negligible contribution to the local stellar mass density relative to the stellar density already in place. To understand whether these galaxies are caught in the act of changing their structural classification properties (i.e. by increasing their central mass concentration, thereby forming their bulge components, or increasing their stellar mass density at large radii, thereby growing the sizes of their disk components) we switch to a diagnostic that is able to trace the change in shape of the stellar mass profiles, i.e. the sSFR profile defined as

\begin{equation}
\mathrm{sSFR}(r) = \Sigma_{\rm SFR}(r)/\Sigma_{\rm M}(r)
\end{equation}
\noindent
where $\Sigma_{\rm SFR}$ is the SFR surface density profile and $\Sigma_{\rm M}$ is the stellar mass surface density profile. With our definition of stellar mass as the integral of the past SFR, sSFR gives directly the inverse of the mass $e$-folding (i.e. roughly the mass-doubling) timescale. 

These sSFR profiles are shown in Figure~\ref{fig:sSFR_profiles}. The top and bottom panels show the sSFR based on the UV and H$\alpha$ SFR, respectively. We show the individual profiles and the stacked profiles in mass bins given by $10.0<\log~M_{\star}/M_{\odot}<11.0$ and $11.0<\log~M_{\star}/M_{\odot}<11.5$. There is a wide range of shapes in the individual profiles, especially in the innermost and outermost regions, but the overall trend is that lower-mass galaxies have on average flat sSFR profiles, and high-mass galaxies have centrally suppressed sSFR profiles. The H$\alpha$ and the UV SFR tracers agree well with each other and reproduce this same result.

Most importantly, these results stand true not only when adopting the radially constant attenuation profiles but also with the newly derived radially varying dust attenuation corrections. Quantitatively, there are of course differences. Most evident is the difference in the lower-mass bin for the UV-based profiles: the radially constant dust correction gives an outward-increasing sSFR profile; introducing the measured radial dependence for the dust correction leads to a flat sSFR profile. A flat sSFR profile is also what is observed in H$\alpha$, which emphasizes the importance of implementing the radial-dependent dust correction to the UV indicators in order to bring the two to agreement. Furthermore, the error bar in the lower right panel of Figure~\ref{fig:sSFR_profiles} indicates the systematic uncertainty on the measured sSFR: the largest contribution comes from the UV (H$\alpha$) to SFR conversion coefficient (variations from the IMF and stellar library), but also incorporates the uncertainty from the dust attenuation estimation. Even in the light of this large and generously estimated uncertainty, the decline of the sSFR toward the central region is evident. Furthermore, the central sSFR values can be considered as upper limits (in particular of the most massive galaxies in our sample) because of the age-dust degeneracy that would lead us to infer high attenuation values for galaxies with prominent, nearly quiescent bulges.

The fact that lower-mass galaxies have flat sSFR profiles and high-mass galaxies have centrally suppressed sSFR profiles, even when radial variations of dust attenuation are accounted for, is a result of significance: it confirms that, below the mass scale of $\sim10^{11}~M_\odot$, SFGs on the SFMS at $z\sim2$ are vigorously doubling their stellar mass at a similar pace in their inner (`bulge') and outer (`disk') regions. In contrast, (at least some of) the most massive of them sustain their high SFRs in their outer disks and host almost-quenched inner `bulges', as already pointed out in \citet{tacchella15_sci}. This agrees well with the observations that, at all epochs, SFGs grow on an almost time-independent correlation between stellar density within the central kiloparsec and total stellar mass (dubbed ``structural main sequence'' by \citealt{barro17}), until they reach a mass of order $\sim10^{11}~M_\odot$ when quenching intervenes \citep{peng10_Cont, tacchella15_sci}.

More in detail, at $\log~M_{\star}/M_{\odot}<11.0$ we measure on average an sSFR value of $<\!\mathrm{sSFR}\!>\approx3~\mathrm{Gyr}^{-1}$, which is consistent with the sSFR value of the SFMS at $z\sim2.2$ (a mass-doubling time of ~0.3 Gyr) at this mass scale \citep[e.g.,][]{rodighiero14, whitaker14, schreiber15}. In contrast, the higher-mass galaxies $\log~M_{\star}/M_{\odot}>11.0$ have on average sSFR values that range from $<\!\mathrm{sSFR}_{\rm out}\!>\approx2~\mathrm{Gyr}^{-1}$ at large radii ($\ga4$ kpc or $\ga1~R_{\rm M}$), again consistent with the SFMS estimates quoted above, to nearly one order of magnitude lower, $<\!\mathrm{sSFR}_{\rm in}\!>\approx0.1-0.4~\mathrm{Gyr}^{-1}$ in the inner bulge regions. So even though the SFR profiles are centrally peaked, the sSFR profiles drop at the centers. This low inner sSFR value implies a mass-doubling time of $\sim2.5-10.0$ Gyr, which is comparable to or larger than the Hubble time at $z\sim2$. The star-formation activity that is taking place in the centers of these galaxies will therefore not significantly increase the central stellar mass density in these systems: their dense bulge components are already in place, as also argued in \citet{tacchella15_sci}. Similarly, \citet{jung17} find evidence for reduced sSFRs in the centers of massive galaxies at $z\sim4$. The direct comparison of the UV- and H$\alpha$-based sSFR profiles shows not only a qualitative but also a quantitative agreement for the values above. 

Our findings for the sSFR profiles also have implications for the evolution of the size-mass relation of galaxies and for identifying the main physical drivers behind this evolution. The fact that at masses below and above $\log~M_{\star}/M_{\odot}\sim11.0$ the radial sSFR profiles are, respectively, flat and outward increasing is overall consistent with studies of the average size growth of the SFG population. In the $\log~M_{\star}/M_{\odot}<11.0$ mass regime, find a slower evolution with cosmic time, and a shallower slope for the mass-size relation at any epoch, relative to the higher-mass population \citep[e.g.,][]{shen03,franx08, van-dokkum08, cassata13, cibinel13a, mosleh11, newman12a, carollo13a, belli14, van-der-wel14a}. 

As shown in \citet{tacchella15_sci}, these galaxies have saturated their stellar mass densities within several kiloparsecs to the values that are observed in the $z=0$ spheroid population of similar mass. The key result that we report in this paper is that obscuration by dust is not responsible for the low sSFR in their cores and thus the quantification of their inner sSFR, which demonstrates that only a negligible amount of mass will be added to their inner regions in their subsequent evolution. This, together with the fact that they are massive and cannot keep forming stars for long in order to avoid dramatically overshooting the highest observed masses of lower-redshift galaxies \citep{renzini09}, implies that these galaxies, which will soon leave the SFMS to reach their final resting place on the quenched sequence, will do so bringing with them, already in place, the high inner stellar densities that we identify with the $\sim10^{11}~M_\odot$ spheroid population in the local universe. Direct quenching from the SFMS to the red and dead cloud should be no surprise at the mass scales that we are discussing. Indeed, the bulk of the spheroid population around $M_{\star}\sim10^{11}M_\odot$ shows all signs of a highly dissipative formation history, i.e. disky isophotes \citep{bender88}, cuspy nuclei \citep{faber97}, outer disks \citep{rix99}, generally a high degree of rotational support \citep{cappellari16}, and steep metallicity gradients \citep{carollo93}. Also at $z\sim1-2$, there is growing evidence that quiescent galaxies are disk-like and rotating \citep{van-der-wel14, chang13, van-de-sande13, newman15, toft17}. We note that this is in contrast with the ultramassive spheroid population at $M_{\star}>>10^{11}M_\odot$: such ultramassive spheroids are very rare and, as already commented in the Introduction (see references there and also \citealt{faisst17_size}), bear the clear signs that dissipationless processes such as dry mergers must characterize their assembly (see \citealt{carollo13a, fagioli16} for extensive discussions on the $M_{\star}\sim10^{11}M_\odot$ mass threshold separating quenched spheroids with dissipative properties from more massive counterparts with dissipationless properties).

The evidence for a direct quenching channel of $\sim 10^{11}~M_\odot$ galaxies from the SFMS to the quenched population supports the picture in which, at any epoch, the universe keeps adding, to the red and dead population, quenched galaxies that inherit the properties -- including the sizes -- of their progenitors on the SFMS \citep{carollo13a, fagioli16}. Since SFG sizes are seen to roughly scale similarly to the host halos with $(1+z)^{-1}$ \citep{oesch10, mosleh11, newman12a, somerville17}, this `progenitor bias' effect \citep{van-dokkum96}, i.e. the addition of larger and larger quenched galaxies to the overall population, becomes therefore a major contribution to the observed size growth of the quenched population with cosmic time. The smaller sizes at any given epoch of quenched galaxies relative to SFGs have been shown to be well explained precisely by the cumulative effects of progenitor bias that piles on the quenched population earlier generations of SFGs \citep{lilly16}, coupled with the fading of the stellar populations in the disks, once also these outer galactic components exhaust their star-formation activity \citep{carollo16}.

\subsection{Evidence for bulge and disk growth on the upper and lower envelope of the SFMS}
\label{subsec:sSFR_dist}

\begin{figure}
\includegraphics[width=\linewidth]{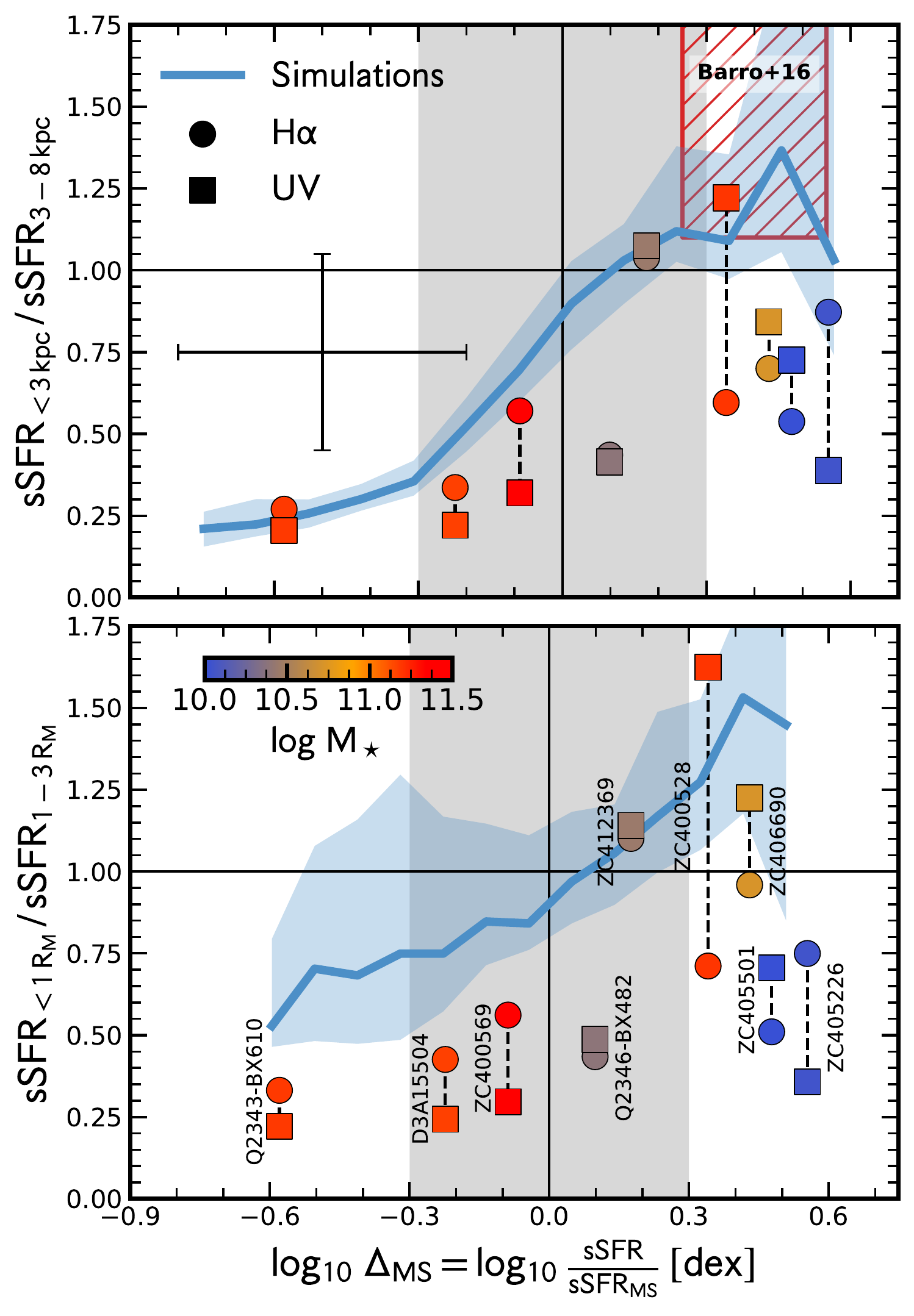} 
\caption{Ratio of the sSFR in the galaxies' centers and outskirts (sSFR$_{\rm in}$/sSFR$_{\rm out}$) as a function of the distance from the SFMS. The top panel shows sSFR$_{\rm in}$/sSFR$_{\rm out}$ in physical units (`in' measured in the $0-3$ kpc range, and `out' in the $3-8$ kpc range); the bottom panel shows the same quantities computed in radial ranges normalized to half-mass radius, $R_{\rm M}$ (respectively, within $0-1~R_{\rm M}$ and $1-3~R_{\rm M}$). The circles and squares show the sSFR estimated from H$\alpha$ ($f=0.70$) and UV, respectively. The color coding shows the total stellar mass. The red hatched region indicates observations by \citet[][see text for details]{barro16}. The error bar in the top panel on the left side indicates the systematic uncertainty (including variations of the IMF). The overall trend for an $\mathrm{sSFR}_{\rm in}\ga\mathrm{sSFR}_{\rm out}$ at the upper envelope of the SFMS, and vice versa at its bottom envelope, is also found in cosmological simulation and there is seen to happen as a result of bulge-forming `compaction' (upper envelope) and disk-forming central depletion (lower envelope) events \citep{tacchella16_MS}.} 
\label{fig:sSFR_in_vs_out}
\end{figure}

With dust-corrected sSFR profiles in hand, we can turn to asking whether these show any difference depending on where the galaxies lie on the SFMS, not simply `along' the sequence, i.e. as a function of stellar mass (discussed before), but `across' the sequence, i.e. above and below the SFMS ridgeline. Numerical simulations make predictions on what happens to a galaxy as it grows its mass along the SFMS, with observational consequences, which we can test with our data. In particular, \citet{tacchella16_MS} find that, in cosmological zoom-in hydrodynamical simulations (i.e. the VELA suite of \citealt{ceverino14_radfeed}), SFGs oscillate up and down the average SFMS, reaching the upper envelope when gas vigorously flows toward the galaxy centers, where it reaches very high densities. This phase, dubbed `compaction' in \citet{zolotov15} \citep[see also][]{tacchella16_profile, tacchella16_MS}, leads to a strong central starburst; it is this starburst that pushes galaxies toward the upper SFMS envelope. The starburst adds substantial stellar mass in the galaxy centers, i.e. to the bulge component. Following the compaction phase, SFGs deplete of gas in the centers owing to the combination of gas consumption into stars and strong outflows from stellar feedback. This indeed halts the compaction phase and push galaxies down toward the lower envelope of the SFMS. 

In particular, we are now going to interpret our results in the framework discussed by \citet{zolotov15} and \citet{tacchella16_profile, tacchella16_MS}. The testable prediction of this scenario is that the spatial distribution of the sSFR within galaxies should correlate with their position on the SFMS: galaxies above (below) the SFMS ridge should have a higher (lower) sSFR in their centers than in their outskirts. 

In Figure~\ref{fig:sSFR_in_vs_out} we plot the ratio of the sSFR in the centers and in the outskirts (sSFR$_{\rm in}$/sSFR$_{\rm out}$) as a function of the distance from the SFMS ($\Delta_{\rm MS}$). The distance from the SFMS is defined as the $\log$-difference of the sSFR of the galaxy to the one of the average SFMS, for which we use a simple linear fit to our sample. The latter agrees well with the SFMS estimates of, e.g., \citet{whitaker14} and \citet{schreiber15}. For the SFR of the SFMS we adopt the total UV+IR SFR that we discuss in Section~\ref{subsec:SFRIR_vs_SFRUV}. To separate the `center' and `outskirts' of the galaxies, we use two different definitions, one in terms of an absolute physical threshold in radius, and the other in terms of radius normalized by the half-mass radius $R_{\rm M}$. We plot in the top panel the inner vs. outer sSFR ratio as a function of radius in physical units in kpc, specifically adopting a 3 kpc threshold to separate inner and outer regions. In the bottom panel we plot instead the same ratio as a function of normalized radius, specifically adopting $1~R_{\rm M}$ as the separating threshold. 

Overall, we observe that galaxies in our sample that lie above the average SFMS ($\Delta_{\rm MS}>0.0$) have a higher or at least comparable sSFR in their centers than in their outskirts ($\mathrm{sSFR}_{\rm in}\ga\mathrm{sSFR}_{\rm out}$); in contrast, we observe the opposite for galaxies below the average SFMS. This result is robust, and our conclusions do not change when using both physical and normalized units, or when varying the boundaries of the definition of `center' and `outskirts'. 

Of course, we must exercise care in commenting on this trend, since our galaxy sample is small, and even smaller are thus the below- and above-SFMS subsamples that we are studying. Our (small) sub-SFMS galaxy sample is furthermore dominated by the most massive galaxies ($M_{\star}\ga10^{11}~M_{\odot}$), which are most probably on the way to being quenched. The sSFR ratio vs. $\Delta_{\rm MS}$ relation is, however, not solely driven by stellar mass. The trend of higher sSFR in the center vs outskirt depends more strongly on the distance from the main-sequence ridgeline than on stellar mass: the Pearson correlation coefficient is $R=0.44~(0.56)$ for UV (H$\alpha$) sSFR ratio versus distance from the main sequence and $R=-0.09~(-0.27)$ for UV (H$\alpha$) sSFR ratio versus stellar mass.

We have highlighted in Section~\ref{subsec:Sample} that our sample traces the SFMS, but we do not probe the massive and dusty SFGs in the upper right corner of the $UVJ$ diagram, which may partially be due to small number statistics in addition to our sample selection criteria. Massive, dusty SFGs and therefore IR-bright galaxies have been recently studied with ALMA by \citet{barro16} and \citet{tadaki17}. \citet{tadaki17} show that their galaxies lie on extrapolation of the \citet{whitaker14} low-mass SFMS; however, we note that they are actually located on the upper envelope of the SFMS when the bending at high masses of the SFMS -- as observed by \citet{whitaker14} -- is taken into account. Furthermore, these galaxies appear to be actively star-forming in their central regions, i.e. the sSFR profiles are rising toward the center. The region in Figure~\ref{fig:sSFR_in_vs_out} occupied by the corresponding profiles of \citet{barro16} is shown as the red hatched area. This is consistent with the framework presented here, where galaxies at the upper envelope of the SFMS are experiencing a dusty nuclear starburst and are doubling their mass quicker in their centers than in their outskirts.

Although our SINS/zC-SINF and the ALMA samples are selected very differently, are small, and are not representative of the whole SFG population at $z\sim2$, a coherent picture emerges when combining the results presented here and in \citet{barro16} and \citet{tadaki17}. These observations are consistent with galaxies growing their inner bulge and outer disk regions as found in the simulations, where they appear to oscillate about the average SFMS in cycles of central gas compaction, which leads to bulge growth, and subsequent central depletion due to feedback from the starburst resulting from the compaction \citep{zolotov15, tacchella16_profile, tacchella16_MS}.

\section{Summary and Concluding Remarks}\label{sec:summary}

As highlighted in the Introduction, the spatially resolved dust attenuation distribution in high-$z$ galaxies still is poorly understood owing to the scarcity of empirical constraints. However, it can have a significant impact not only on the inferred star formation distribution of galaxies but also on the measurement of sizes and shapes, the estimation of the stellar mass surface density, the identification of star forming clumps, and the conversion of the H$\alpha$ luminosity to the gas surface density. In this paper, we have therefore combined \textit{HST} rest-frame optical, NUV, and FUV imaging with \textit{VLT} SINFONI AO integral field spectroscopy in 10 massive SFMS galaxies at $z\sim2.2$, all resolved on scales of $\sim1$ kpc, to derive 2D distributions of dust attenuation from the UV slope $\beta$, and thus dust-corrected SFRs as well as dust-corrected UV and H$\alpha$ SFR profiles. In particular, we assume that SFRs derived from the attenuation-corrected UV luminosity are reliable in all but possibly a minority of cases with extremely high attenuation. Moreover, we also assume that the UV attenuation A$_{\rm UV}$ can be used to derive the attenuation at H$\alpha$ (A$_{\rm H\alpha}$), hence allowing us to construct space-resolved SFR maps from locally corrected H$\alpha$ flux maps.

The radial profiles of dust attenuation A$_{\rm V}$ vary substantially from galaxy to galaxy, but overall they show a general trend to increase toward the galaxy centers, as also found in other samples \citep{wuyts12, hemmati15, nelson16_balmer, wang17_dust}. In our work we find only a very weak dependence of the A$_{\rm V}$ profile's shape on the total stellar mass, with more massive galaxies showing a slightly higher attenuation in their centers than lower-mass galaxies. Our sample displays a relatively shallow gradient, leading to a significant amount of dust attenuation out to large galactocentric distances: at $\sim10$ kpc, the average A$_{\rm V}$ value is $\approx0.6$ mag.

We use these radial profiles of attenuation to correct the radial profiles of SFR and sSFR, which we then interpret more specifically in the framework of our previous work, with a discussion of implications on relative disk versus bulge growth. In particular, we find an important trend for the sSFR profiles with stellar mass in our sample. Our galaxies with masses below $\sim10^{11}~M_{\odot}$ are on average doubling their stellar mass at a similar pace in their inner and outer regions, indicating that they are synchronously growing their `bulge' and `disk' regions. This agrees well with the observations that, at all epochs, SFGs on the SFMS grow on an almost linear correlation between stellar density within the central kpc ($\Sigma_1$) and total stellar mass ($M_{\star}$) -- until, as further discussed below, they reach a mass of order $\sim10^{11}~M_\odot$ at which they start their transition toward the quenched population (see also \citealt{saracco12, tacchella15_sci, barro17, tacchella17_S1}).  

At higher masses, at and above $\sim10^{11}~M_{\odot}$, the galaxies in our sample show a reduced star-formation activity in their centers with respect to their outskirts. The mass-doubling time in the center of these galaxies exceeds the Hubble time by about a factor of two ($\mathrm{sSFR}^{-1}>t_{\rm H}$), in agreement with the main result of \citet{tacchella15_sci}. Important to notice is that the central sSFR values can be considered as upper limits because of the age-dust degeneracy that would lead us to infer high attenuation values for galaxies with prominent, nearly quiescent bulges. We interpreted such centrally depleted sSFR profiles of massive galaxies as a signature of `inside-out quenching'., but it is also consistent with `inside-out growth' of massive galaxies (as, e.g., in \citealt{nelson16_insideout, lilly16}). It is important to emphasize that the observed central depression in sSFR in our most massive galaxies rests on the assumption that the UV slope $\beta$ is entirely due to reddening, as opposed to the prevalence of an old, passively evolving population. Therefore, our central $\Sigma_{\rm SFR}$ can be regarded as upper limits, due to the mentioned age-reddening degeneracy. In any event, these centrally depressed sSFRs can help the consistency of the evolution of the SFMS with observed  evolution of the cosmic SFR density \citep{renzini16}.

By combining our observations with those by \citet{barro16} and \citet{tadaki17}, we find that galaxies above and below the average SFMS relation appear to have respectively centrally enhanced and centrally suppressed sSFRs, respectively. Similar trends have been recently found in the local universe \citep{ellison18, belfiore17_quenching}. This is consistent with a picture suggested by numerical simulations where galaxies oscillate up and down with cosmic time relative to the SFMS ridgeline, alternating phases of gas `compaction' and bulge growth, followed by gas depletion and suppressed star formation in the central regions \citep{tacchella16_profile, tacchella16_MS}.
 
In concluding, we emphasize strengths and weaknesses of the present investigation. On the positive side, rest-frame FUV, NUV, and optical \textit{HST} imaging, together with resolved H$\alpha$ spectroscopy at a similar kiloparsec resolution, gives us a unique dataset to learn about the stellar mass distributions within galaxies and the locally related dust-corrected SFR indicators from two independent diagnostics. On the side of caveats, it is important to remember the by-necessity simplified assumptions on dust distribution and properties that we discussed in the Introduction and in Section \ref{subsec:caution}. Also, our sample is limited to only 10 galaxies, which does not probe the whole parameter space of the massive galaxy population at $z\sim2$ (Figure~\ref{fig:Sample}). In particular, our sample does not include the most heavily obscured, massive galaxies. We also point out that, in order to obtain the attenuation values A$_{\rm V}$ and A$_{\rm H\alpha}$, we have used average relations that are well suited for application to large statistical studies of SFGs, but should be used with caution on a galaxy-by-galaxy basis and on spatially resolved scales. Finally, while the comparison with IR-based SFR disfavors the presence of regions of total obscurations in our sample, full proof of this will need high-resolution imaging of the molecular gas and of the continuum dust emission. Continued efforts combining maps of UV, IR/submillimeter, and hydrogen recombination lines from JWST and NOEMA/ALMA will be very important to make further progress on the key issue of galaxy mass buildup.


\acknowledgments
We thank the first referee for helping us to improve the discussion of our sample and the second referee for significantly improving the discussion of our results. We thank Marc Rafeski for providing us with his improved dark calibrations for the reduction of the WFC3/UVIS data \citep{rafelski15}, and Sandy Faber for inspiring discussions. ST also thanks also Jonathan Freundlich, Nicholas Lee, Gabriele Pezzulli, Romain Teyssier, Benny Trakhtenbrot, and Weichen Wang for useful suggestions. We acknowledge generous support by the Swiss National Science Foundation. This work was partly supported by the grants ISF 124/12, I-CORE Program of the PBC/ISF 1829/12, BSF 2014-273, PICS 2015-18, and NSF AST-1405962. This research made use of NASA's Astrophysics Data System (ADS), the arXiv.org preprint server, the Python plotting library \texttt{matplotlib} \citep{hunter07}, and \texttt{astropy}, a community-developed core Python package for Astronomy \citep{astropycollaboration13}.


{\it Facility:} \facility{HST (WFC3, ACS)}, \facility{VLT (SINFONI)} \\



\appendix
\section{Dust Attenuation Parameters from (FUV$-$NUV) Color}\label{App:color_dust}
\subsection{Conversion}

\begin{figure}
\centering
\includegraphics[width=\linewidth]{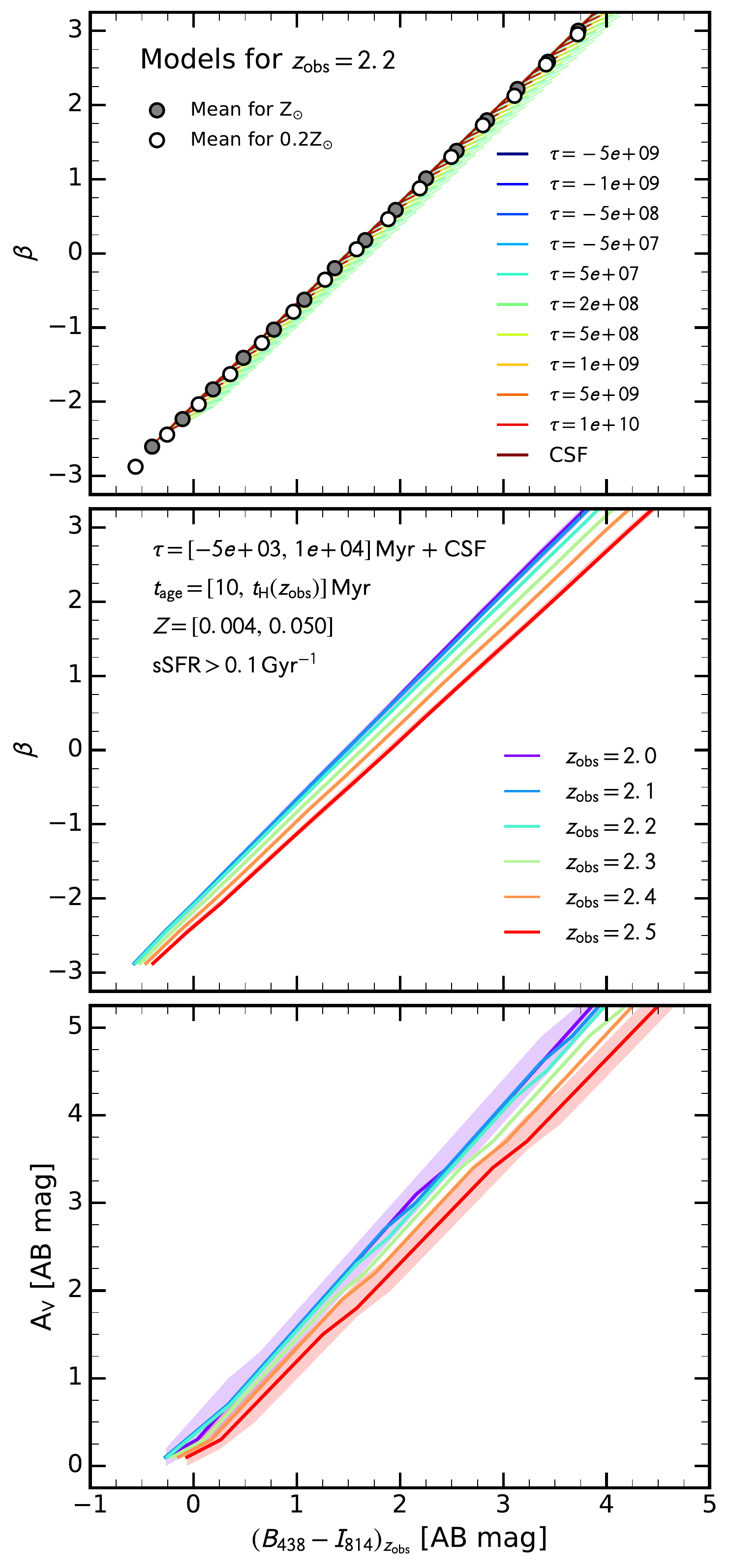} 
\caption{Relations between dust attenuation (UV continuum slope $\beta$ and total attenuation at $V$ A$_{\rm V}$) and observed color $B_{428}-I_{814}$ (rest-frame FUV-NUV). Top panel: relation between $\beta$ and $B_{428}-I_{814}$ for a large grid of different SED models at a fixed redshift $z_{\rm obs}=2.2$ and for a star-forming population ($\mathrm{sSFR} > 0.1~\mathrm{Gyr}^{-1}$). The various curves plotted in colors are computed from \citet{bruzual03} models with solar metallicity and a \citet{chabrier03} IMF. Different colors are used for different SFHs: a suite of exponentially declining and increasing SFRs with different $e$-folding timescales, and constant SFR. Age increases along each model curve from blue to red $B_{428}-I_{814}$ colors and low to high $\beta$. The gray filled and white filled circles show the mean relationship derived from solar and 1/5 solar metallicity models. Middle and bottom panels: mean dependence of $\beta$ and A$_{\rm V}$ on $B_{428}-I_{814}$ at different redshifts ($z_{\rm obs}=2.0-2.5$). Overall, the relations between dust attenuation and the observed $B_{428}-I_{814}$ are tight and have a weak redshift dependence. Therefore, we use the $B_{428}-I_{814}$ to estimate the dust attenuation as given in Equations~\ref{eq:color_beta} and \ref{eq:color_AV} (best-fitting relations).} 
\label{fig:App_color_dust}
\end{figure}

We derive the (FUV$-$NUV)$-\mathrm{A_{\rm V}}$ and (FUV$-$NUV)$-\beta$ conversions in Figure~\ref{fig:App_color_dust} by generating a set of model SEDs from \citet{bruzual03}, for six different metallicities (Z=0.0001, 0.0004, 0.004, 0.008, 0.02, and 0.05) and three different SFHs (constant, exponentially rising, and exponentially decreasing). The timescales used for the rising and declining SFHs are respectively $\tau= [-5000,-1000,-500,-50,50,250,500,1000,5000,10000]$ Myr, respectively. Note that the negative sign indicates rising SFHs. The stellar age is defined as the age since the onset of star formation. We consider a minimum age of 10 Myr and a maximum age of 3.5 Gyr (age of the universe at $z\sim2$). We assume that the dust attenuation is described by the Calzetti dust attenuation curve \citep{calzetti00}, with $\mathrm{A_{\rm V}}$ varying from 0.0 to 6.0, in steps of 0.05 between 0.0 and 4.0, in steps of 0.25 between 4.0 and 5, and in steps of 0.5 between 5.0 and 6.0. The intergalactic medium (IGM) is treated using the \citet{madau95} prescription. Redshifts vary between 2.0 and 2.5 in steps of 0.1. 

As shown in the top panel of Figure~\ref{fig:App_color_dust}, at a fixed redshift we find a tight correlation between the observed $B_{428}-I_{814}$ color (approximately corresponding to the rest-frame FUV-NUV color) and $\mathrm{A_{\rm V}}$, with a small  but significant dependence on redshift (see middle and bottom panels). These correlations are well approximated with

\begin{equation} 
\begin{split} 
\mathrm{A_{\rm V}} = & (2.36\pm0.11)+(2.11\pm0.01)\times(B-I)_{z_{\rm obs}} \\ 
 & -(4.11\pm0.40)\times\log(1+z_{\rm obs}) \\ 
 & -(1.78\pm0.05)\times \log(1+z_{\rm obs}) \times (B-I)_{z_{\rm obs}},
 \label{eq:color_AV}
\end{split}
\end{equation}

\noindent where $(B-I)_{z_{\rm obs}}$ is the observed $(B-I)$ color in mag and $z_{\rm obs}$ is the redshift of the observed galaxy. 

For $z_{\rm obs}=2.2$, we get

\begin{equation} 
\mathrm{A_{\rm V}} = (0.28\pm0.41)+(0.77\pm0.05)\times(B-I)_{z_{\rm obs}=2.2}
 \label{eq:color_AV_typical}
\end{equation}

\noindent similar to the expression suggested by \citet{meurer99}.

For each model SED, we calculated the UV slope ($\beta$) by fitting the flux of the SED model as a function of wavelength, using only those flux points that lie within the 10 continuum windows given in \citet{calzetti94}. The typical formal uncertainty in $\beta$, when using the 10 aforementioned windows, is $\Delta\beta\approx0.1$. 

We find a tight correlation between the (FUV$-$NUV) color and $\beta$, similar to above:

\begin{equation} 
\begin{split} 
 \beta = & (0.41\pm0.08)+(2.28\pm0.01)\times(B-I)_{z_{\rm obs}} \\ 
 & -(5.03\pm0.30)\times\log(1+z_{\rm obs}) \\ 
 & -(1.86\pm0.04)\times \log(1+z_{\rm obs}) \times (B-I)_{z_{\rm obs}}.
 \label{eq:color_beta}
\end{split}
\end{equation}

For $z_{\rm obs}=2.2$, we get

\begin{equation} 
 \beta = (-2.13\pm0.31)+(1.34\pm0.04)\times(B-I)_{z_{\rm obs}=2.2}.
 \label{eq:color_beta_typical}
\end{equation}

\subsection{Stellar Population Parameters affecting the UV Continuum Slope}

\begin{figure}
\centering
\includegraphics[width=\linewidth]{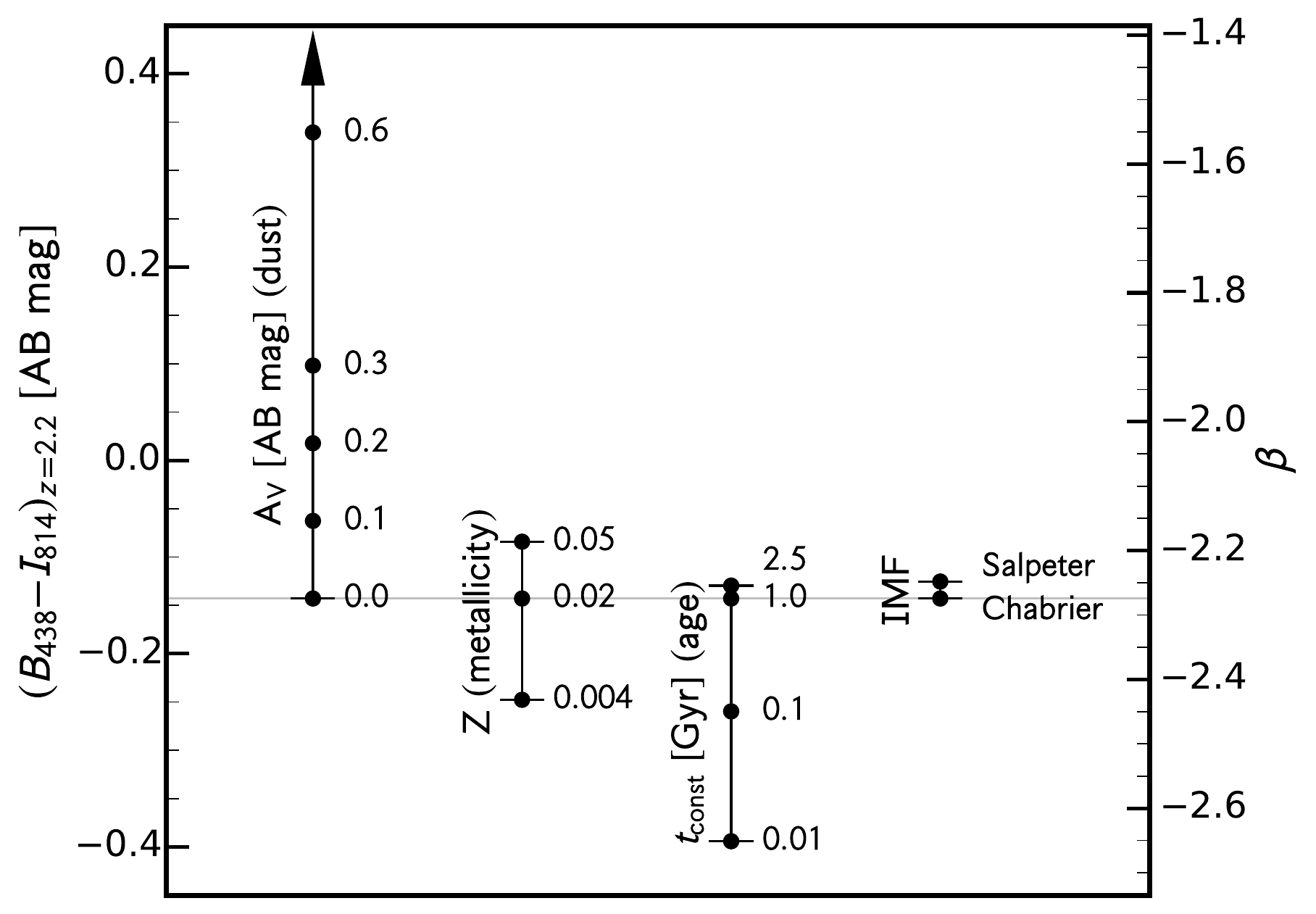} 
\caption{Overview of the factors influencing the observed $(B-I)$ color (rest-frame UV continuum slope). The horizontal line denotes the $(B-I)$ color inferred from the \citet {bruzual03} model assuming the default scenario: 1 Gyr previous continuous star formation, solar metallicity ($Z=0.02$), a \citet{chabrier03} IMF, and no dust. Variations in the metallicity, age, and IMF can affect the $(B-I)$ color by up to 0.3 mag, which translates into change in the inferred dust attenuation parameter of $\la0.3$ mag. } 
\label{fig:App_color_SP}
\end{figure}

\begin{figure}
\centering
\includegraphics[width=\linewidth]{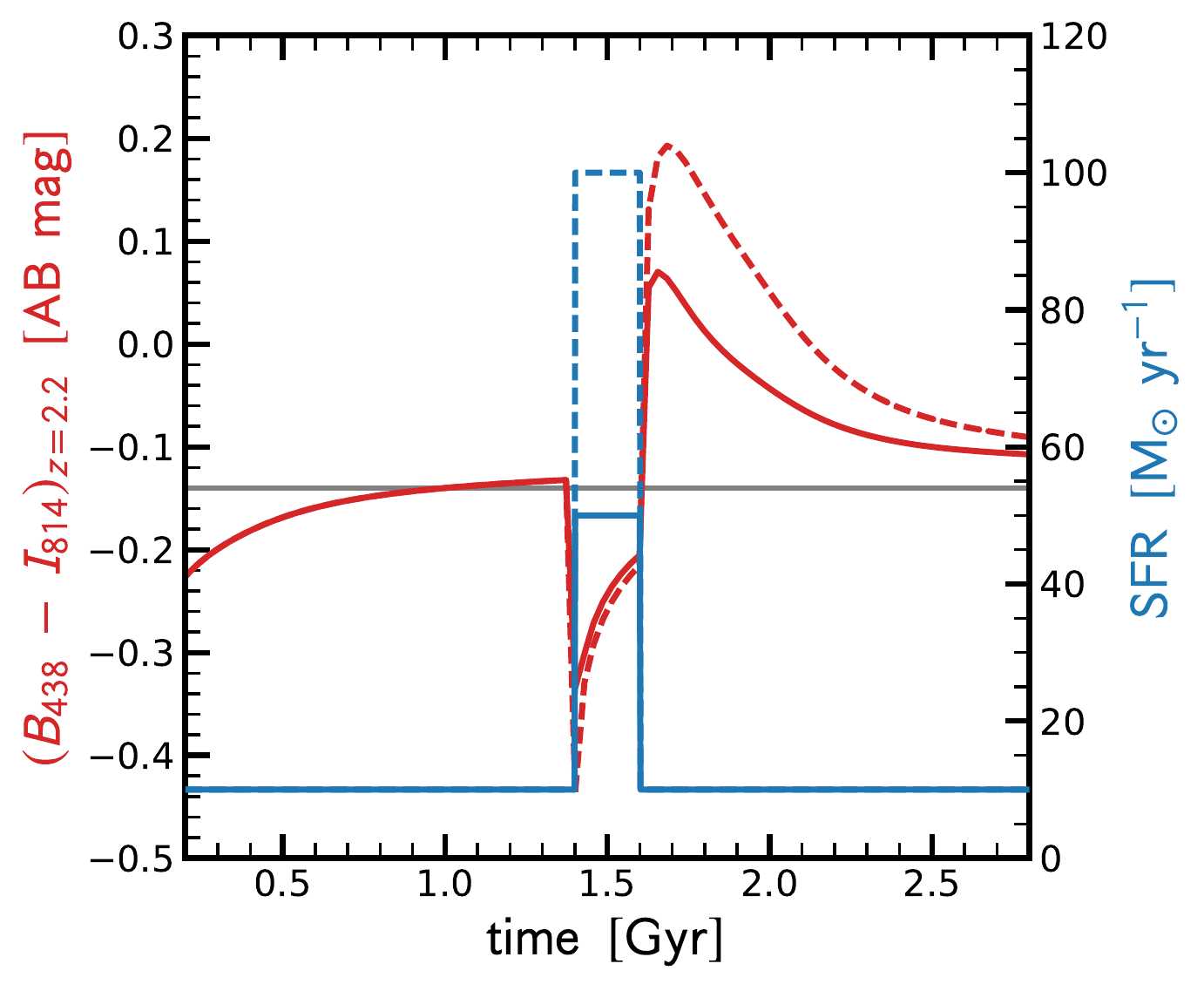} 
\caption{Effect of a single burst on the observed $(B-I)$ color (rest-frame UV continuum slope). The blue lines indicate the SFR as a function of time (right axis), while the red lines show the associated $(B-I)$ color variation as a function of time. The solid and dashed lines highlight the scenarios with a maximal SFR of the burst of $\mathrm{SFR}_{\rm burst}=50~M_{\odot}/\mathrm{yr}$ and $100~M_{\odot}/\mathrm{yr}$, respectively. The horizontal gray line denotes the $(B-I)$ color inferred from the \citet{bruzual03} model assuming the default scenario, as already shown in Figure~\ref{fig:App_color_SP}. At the onset of the burst, the $(B-I)$ color gets bluer by about 0.2 mag. On the other hand, the end of the burst, the $(B-I)$ color reddens significantly by about 0.3 mag since the hottest / bluest stars leave the stellar main sequence first. } 
\label{fig:App_color_burst}
\end{figure}

In this section we investigate the effect of varying the stellar population properties on the (FUV$-$NUV) color \citep[see also, e.g.,][]{cortese08, wilkins11}. In a first step, to compare the different effects, we take a reference model about which we consider deviations in the amount of dust, the previous SFH, metallicity, and IMF. This reference model assumes 1 Gyr continuous star formation, solar metallicity (Z = 0.02), a \citet{chabrier03} IMF, and no dust and is constructed using the \citet{bruzual03} population synthesis model. As shown in Figure~\ref{fig:App_color_SP} (see also \citealt{wilkins11} for a similar figure), this scenario suggests a rest-frame $(\mathrm{FUV}-\mathrm{NUV})$ color of $-0.14~\mathrm{mag}$.

By far the largest potential effect on the UV color of SFGs is the presence of dust: changing $\mathrm{A_{\rm V}}=0.0\rightarrow0.5~\mathrm{mag}$ results in $\Delta(\mathrm{FUV}-\mathrm{NUV})\approx 0.4~\mathrm{mag}$. The strong wavelength dependence of typical \citep[e.g.,][]{calzetti00} reddening curves results in greater extinction in the UV (relative to the optical or near IR) and the reddening of UV colors. Therefore, reddening of the UV colors provides a good diagnostic of the magnitude of the dust attenuation. Next, we vary the second-order stellar population parameter in order to constrain the typical uncertainty that is related to estimating A$_{\rm V}$ from $(\mathrm{FUV}-\mathrm{NUV})$ color.

Variation in the SFH influences the $(\mathrm{FUV}-\mathrm{NUV})$ color since, after prolonged periods of star formation, some fraction of the most massive stars will have evolved off the stellar main sequence, which reduces the relative contribution of these stars to the UV continuum, resulting in a redder color. For example, for a given total stellar mass, reducing the age of a constantly star-forming population by a factor of 10 (i.e., form 1 Gyr to 100 Myr), makes a color by $\Delta(\mathrm{FUV}-\mathrm{NUV})\approx-0.1~\mathrm{mag}$ bluer. Lowering the metallicity from $Z=0.02\rightarrow0.004$ results in $\Delta(\mathrm{FUV}-\mathrm{NUV})\approx-0.1~\mathrm{mag}$. Finally, changing the IMF from \citet{chabrier03} to \citet{salpeter55} makes the color redder by a small amount ($\Delta(\mathrm{FUV}-\mathrm{NUV})=0.01~\mathrm{mag}$) because \citet{salpeter55} IMF is more bottom-heavy. 

In a second step, we quantify the variation of the $(B-I)$ color during a burst of star formation. In detail, the SFR is kept constant at $10~M_{\odot}/\mathrm{yr}$ for 3 Gyr. We superimpose a burst of star formation at time 1.4 Gyr with duration of 200 Myr. We analyze two scenarios: in the first one, the peak SFR during the burst phase is $\mathrm{SFR}_{\rm burst}=50~M_{\odot}/\mathrm{yr}$, while in the second one it is $SFR_{\rm burst}=100~M_{\odot}/\mathrm{yr}$. The SFHs are plotted as blue lines in Figure~\ref{fig:App_color_burst}.

Throughout this variation of the SFR, we compute the associated colors with the same assumptions as for our default scenario, i.e. solar metallicity ($Z=0.02$), a \citet{chabrier03} IMF, and no dust. The obtained colors are shown in red in Figure~\ref{fig:App_color_burst}. After 1 Gyr, the $(B-I)$ color reaches a value of $-0.14$ AB mag, consistent with our default scenario mentioned above. At the onset of the burst, the $(B-I)$ color gets bluer to a value of $-0.33$ AB mag ($-0.43$ AB mag) for $\mathrm{SFR}_{\rm burst}=50~M_{\odot}/\mathrm{yr}$ ($\mathrm{SFR}_{\rm burst}=100~M_{\odot}/\mathrm{yr}$). While the SFR is constant during the burst phase, the $(B-I)$ color exponentially reddens again. Shortly ($\la10$ Myr) after the end of the burst phase, the $(B-I)$ color reddens significantly to 0.07 AB mag (0.19 AB mag) since the bluest and therefore most massive stars leave the stellar main sequence first. The $(B-I)$ color then gets bluer again.

Summarizing qualitatively, a significant increase in the SFR leads to a bluer $(B-I)$ color, while a reduction in SFR leads to a redder $(B-I)$ color, assuming constant metallicity and dust content. Quantitatively, we find that a single burst can lead to changes in the $(B-I)$ color of about 0.3 mag. Hence, for a given $(B-I)$ color ($\mathrm{FUV}-\mathrm{NUV}$ color) we estimate the typical uncertainty on A$_{\rm V}$ due to stellar population differences to be at most $\sim0.4$ mag.

\section{Reliability of SFR from IR}\label{App:IR}

As mentioned in Section~\ref{subsec:SFRIR_vs_SFRUV}, there is the danger of source confusion when using the IR to estimate the SFR. To minimize this problem at the longest wavelengths (i.e. $\lambda \ga 100~\mu\mathrm{m}$), the 24 $\mu$m source detections are used as priors. We check in this section the reliability of the SFR$_{\rm IR + UV}$ for ZC400528 and ZC407306, since both of these galaxies have SFR$_{\rm IR + UV}$ values that are significantly larger than dust-corrected SFR$_{\rm UV}$ values.

\begin{figure}
\centering
\includegraphics[width=\linewidth]{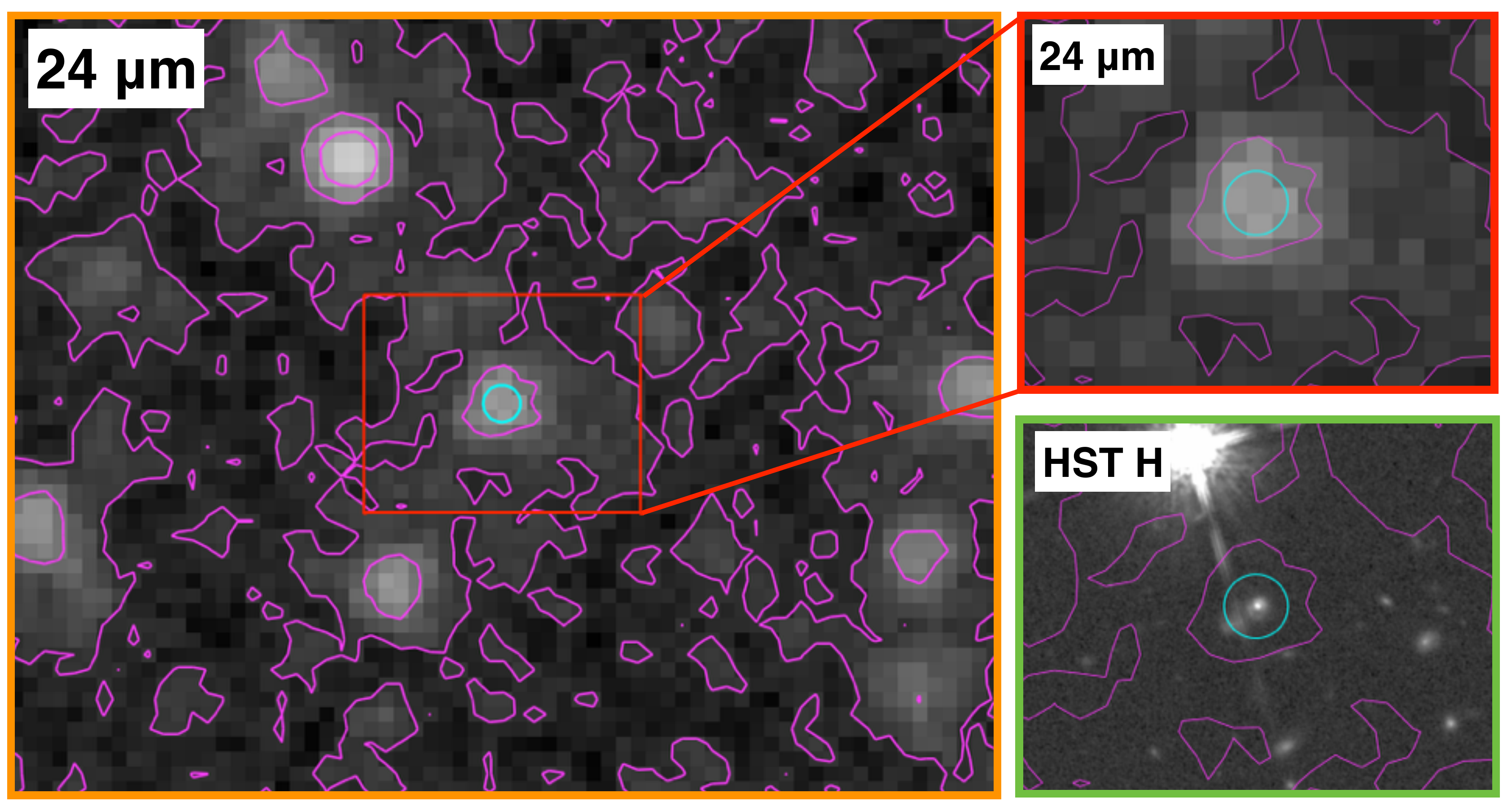}
\caption{Comparison of the \textit{Spitzer} MIPS 24 $\mu$m image with the \textit{HST} WFC3 $H$-band image for galaxy ZC400528. North is up in all panels. The left panel shows a $1.3\arcmin \times 1.0 \arcmin$ cutout of the 24 $\mu$m image. The top right panel shows a zoom-in (approximately $0.4\arcmin \times 0.3 \arcmin$) on the 24 $\mu$m image, which is marked with a red box in the left panel. The bottom right panel shows the same region of the \textit{HST} $H$-band image. In all panels, the magenta lines indicate the contours of the 24 $\mu$m flux map, and the cyan circle shows an aperture of 3\arcsec, centered on ZC400528. There is a close (projected distance is about $1.\arcsec54\approx13$ kpc) neighbor in the southeastern region of ZC400528 that contributes to the 24 $\mu$m and therefore also to the 100 $\mu$m flux, making the SFR$_{\rm 100\mu m +UV}$ difficult to interpret. } 
\label{fig:App_ZC400528}
\end{figure}

\begin{figure}
\centering
\includegraphics[width=\linewidth]{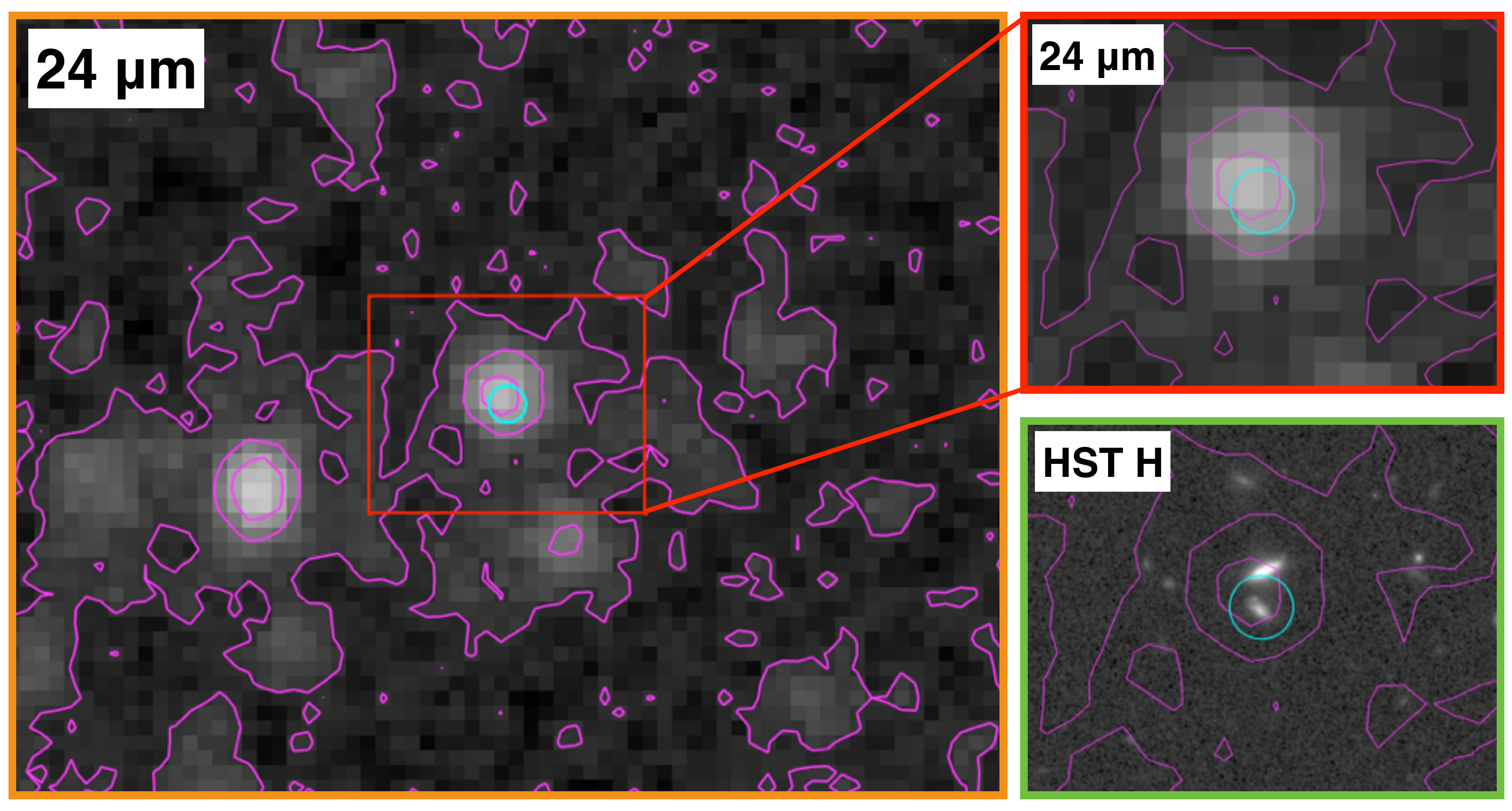}
\caption{Same as Figure~\ref{fig:App_ZC400528}, but for galaxy ZC407302. The 24 $\mu$m detection clearly encompasses ZC407302 and the northern, low-$z$ counterpart seen in the HST image, which contaminates the $24\mu m$ flux and therefore makes the SFR$_{\rm 24\mu m +UV}$ unreliable.} 
\label{fig:App_ZC407302}
\end{figure}

In Figures~\ref{fig:App_ZC400528} and \ref{fig:App_ZC407302} we compare the \textit{Spitzer} MIPS 24 $\mu$m with the \textit{HST} WFC3 $H$-band images for ZC400528 and ZC407302, respectively. ZC407302 has a bright, low-$z$ neighbor that boosts the IR flux and therefore makes the IR photometry of this galaxy unreliable. Hence, throughout the paper, we have used as the fiducial SFR the dust-corrected UV SFR estimate. ZC400528 has also a neighboring galaxy, though this counterpart is much fainter. In addition, ZC400528 shows a nuclear outflow, [\ion{N}{2}]/H$\alpha\approx0.6$, and the 1.4 GHz VLA radio data imply an SFR$\approx790~M_{\odot}~\mathrm{yr}^ {-1}$, all consistent with some AGN activity \citep{forster-schreiber14}. However, it is unclear how much an AGN could contribute to the IR and radio measurements. Therefore, for this galaxy, we use as the fiducial SFR the UV+IR SFR estimate, though further follow-ups are needed, i.e., with spatially resolved dust continuum measurements.

\end{document}